\documentclass[fleqn,10pt]{wlscirep}
\usepackage[utf8]{inputenc}
\usepackage[T1]{fontenc}
\usepackage{epsfig}
\usepackage{amsfonts}
\usepackage{comment}
\usepackage{graphicx}
\usepackage{bm}
\usepackage{amssymb}
\usepackage{color}
\usepackage{float}
\usepackage{amsmath}
\usepackage{amstext}
\usepackage{latexsym}
\usepackage{enumitem}
\usepackage{scalerel}
\usepackage{soul}
\usepackage{xcolor}
\usepackage[normalem]{ulem}
\usepackage{pifont}
\usepackage{braket}

\title{Chaos controlled and disorder driven phase transitions induced by breaking permutation symmetry}

\author[1]{Manju C}
\author[2]{Arul Lakshminarayan}
\author[3,*]{Uma Divakaran}
\affil[1]{Department of Physics, Indian Institute of Technology Palakkad, Kerala, India 678623}
\affil[2]{Department of Physics, \& Center for Quantum Information, Computation and Communication, Indian Institute of Technology Madras, Chennai, India~600036}
\affil[3]{Department of Physics, Indian Institute of Technology Palakkad, Kerala, India 678623}

\affil[*]{uma@iitpkd.ac.in}


\begin{abstract}
The effects of disorder and chaos on quantum many-body systems can be superficially similar, yet their interplay has not been sufficiently explored. This work finds a continuous phase transition when disorder breaks permutation symmetry,  with details of the transition being controlled by the degree of chaos in the clean limit. The system changes from an area law entangled phase in the permutation symmetric subspace where collective variables exist to volume law entanglement in the full Hilbert space, beyond a critical strength of the disorder. 
This has potential implications for general many body physics, as well as technologies such as transmon qubits.

\end{abstract}
\begin{document}

\maketitle
\section{Introduction}
Deterministic chaos, originating in the lack of sufficient number of conserved quantities, gives rise to ergodic and mixing dynamics with a
positive Lyapunov exponent in the classical case \cite{tabor1989chaos, lichtenberg2013regular, Ott_2002}
, and in the quantum case results in states with random matrix properties, thermalization and volume law entanglement \cite{gutzwiller1991chaos, Stockmann_1999, mehta2004random, Sandrobook, Rigol_polkonikov, kaufman2016quantum, Bohigas, srednicki1994chaos}. 
Disorder, depending on its relative strength on the other hand, can lead to Griffiths phases \cite{griffiths69}, Anderson \cite{anderson58} or many-body localization (MBL) \cite{basko06,oganesyan_pal}, and can also result in thermalization and delocalization via breaking symmetries. For example, the extensively investigated Sachdev-Ye-Kitaev (SYK) model, related to the physics of quantum black holes, is one where randomized all-to-all couplings results in maximal quantum chaos \cite{SYK_PRL_1993, kitaev2015simple, maldacena2016remarks}.
Deterministic chaos can sometimes be modeled as being pseudo-random and insights from disorder physics can be applied as in the well-known connection between dynamical and Anderson localization \cite{grempel84}.

Arrays of superconducting Josephson junction based transmons, which are nonlinear oscillators, are currently the most successful implementations of quantum computers \cite{Devoret, PRL_Xmon, transmon_latest}. 
Disorder in the energies of the individual qubit transmons is a necessary evil in the presence of inter-transmon couplings. A detailed study of the consequences for these devices comes with a warning of 
being ``dangerously close to a phase of uncontrollable chaotic fluctuations" \cite{Berke_2022}. 
It has also long been appreciated that dynamical chaos  has serious effects on practical implementation of quantum many-body systems, as it can increase entanglement to random state values, and small subsystems face effective decoherence, thus questioning the reliability of quantum information processing  \cite{zurek2006decoherence, trust_quantum_simulators, bandyopadhyay2002testing,gong2003intrinsic, flambaum2000time, song2001quantum, georgeot2000emergence, braun2002quantum, madhok2018quantum, chaos_and_computers, piga2019quantum, LMG_PRR, papparaldi_bridging}.
For example, the 
presence of random individual qubit energies along with residual short range interactions can make the system chaotic under certain limits that compromises the quantum computer losing its ability to perform well \cite{georgeot2000emergence}.

Classically it was also found that disorder can  ``tame" chaos, wherein nonlinear oscillators get synchronized from a chaotic phase in the presence of disorder \cite{taming_disorder}. 
Similarly, many-body localization takes the system from an ergodic phase to a localized one by increasing disorder \cite{Vonneumannscaling,  MBL_pal, oganesyan_pal, MBL_Luitz}.
It is thus desirable to study the effects of disorder in systems which have a well-established regular $\rightarrow$ chaotic route  in the semiclassical or mean-field limit. Here, the transitions in physical properties whose origins lie in dynamical chaos and those from disorder are both at play together.

Towards this end, we investigate the dynamics of $N$ particles originally restricted to the small $O(N)$ permutation symmetric subspace (PSS), but  that 
can have various dynamical regimes, from near-integrable to intermediate and fully chaotic.  The disorder breaks the permutation symmetry and results in a  phase transition into the full exponentially large, $O(\exp(N))$ Hilbert space. Interestingly, the extent of chaos in the ``clean" symmetric space
is found to dictate the critical strength of the disorder that ensures the dynamics has penetrated to the full Hilbert space (FHS), in the large $N$ limit.
The transition has many possible order parameters. The block entanglement changes from area law $ \sim \ln N$, characteristic of permutation symmetric states \cite{tripartitescrambling} to volume law $\sim N$. Collective observables such as the total angular momentum changes from $\sim N$ to $\sim \sqrt{N}$. 
This transition thus shows similarities to a superradiant to radiant phases of collective atomic states in a cavity \cite{scully_single_photon, J.physicssuperradiant, rubies2022superradiance}. 
Recent studies have explored the effect of introducing non-uniform initial states on the dynamics of a permutation symmetric Hamiltonian \cite{DHS_23, iemini2024dynamics}. In Ref.[{\citenum{DHS_23}}], a similar $\sqrt{N}$ behavior of total angular momentum is observed where the  dynamics penetrate into the ``deep Hilbert space" having certain properties, when the initial state is not from PSS.

\begin{figure}[h] 
\centering
    \includegraphics[width=0.8\linewidth]{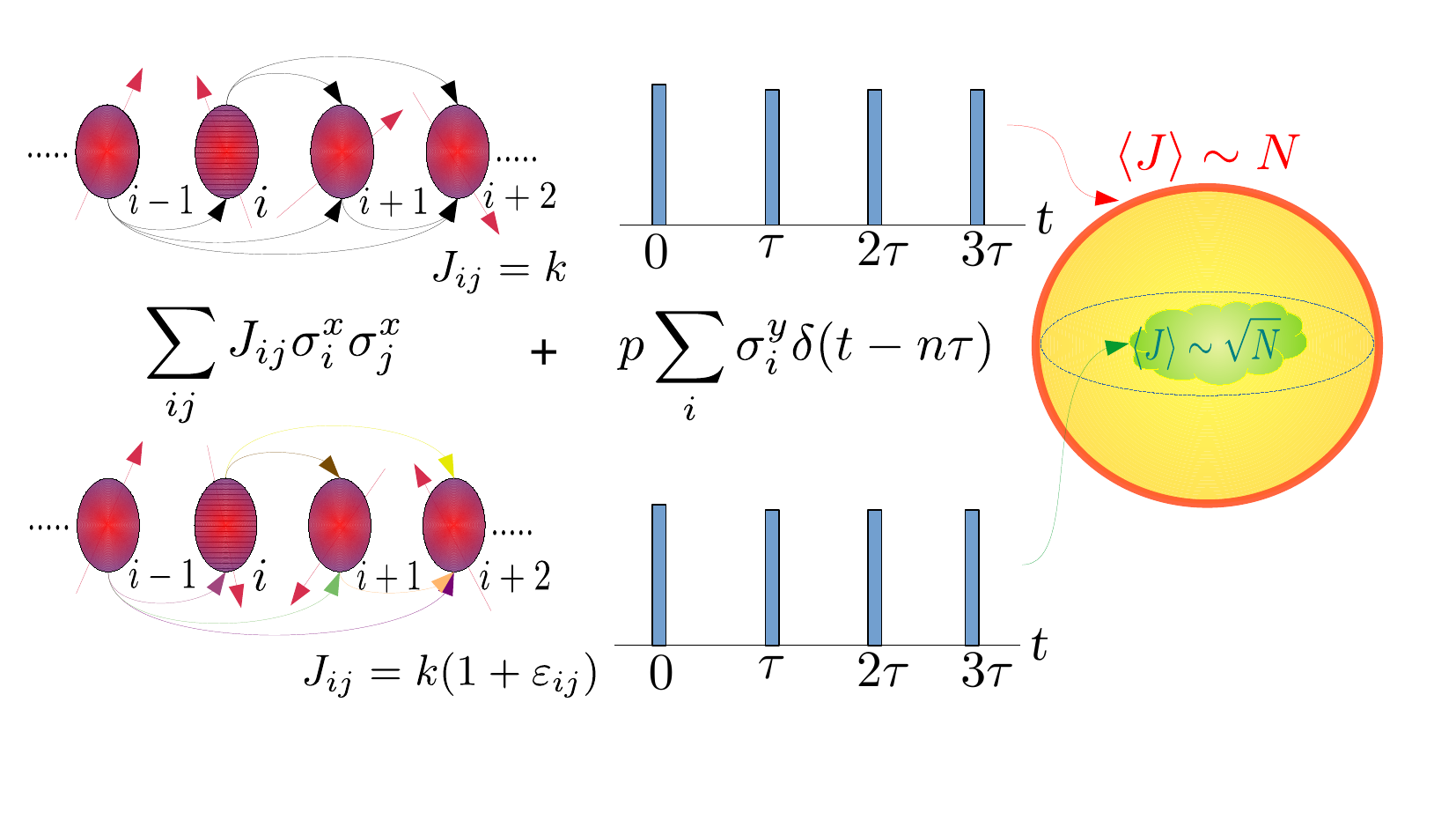} 
    \caption{Schematic of the dynamics: Top row corresponds to the disorder free system where interactions connecting different spins have same magnitude $J_{ij}=k$. The second term of the Hamiltonian is the periodic kicks along $y-$direction in uniform magnetic fields. Bottom row corresponds to a disordered system where $J_{ij}$ are random represented by different colors.
    The angular momentum sphere on the right is a caricature of states with $\langle J \rangle$ for the disorder free system restricted on the surface of this sphere, whereas for the disordered system it is pushed inside the sphere where $\langle J \rangle \sim \sqrt{N}$. 
    }
    \label{fig_schematic}
  \end{figure}

\section{The model}
A convenient model to study the interplay of chaos and disorder is provided by the kicked top which is a textbook example of quantum chaos \cite{haake1987classical, PhysRevA.45.3646, Haakebook}, consisting of one large spin of angular momentum quantum number $j=N/2$, that can be considered as the collective dynamics of $N$ spin $1/2$ particles
within the permutation symmetric subspace \cite{concurrence, madhok_quantum_discord}. Several experiments have fruitfully implemented this to understand the effects of chaos on entanglement \cite{chaudhury_nature_husimi,Neill_2016, NMRstudiesin2qubit}.

A disordered version of the kicked top  can be  used to investigate 
the effects of breaking permutation symmetry.
The specific  Hamiltonian  that we study is given by
\begin{equation}
	H=\frac{k}{2 N \tau}\sum_{\ell< \ell'=1}^{N}(1+\epsilon_{\ell \ell'})\sigma_{\ell}^{x}\sigma_{\ell'}^{x}  +\frac{p}{2}\sum_{n=-\infty }^{\infty }\sum_{\ell=1}^{N}\sigma_{\ell}^{y}\delta (t-n\tau),
\label{eq_spin}
\end{equation}
where, $\sigma_{\ell}^\nu$ with $\nu=x~,y,~z$ are the Pauli 
matrices at site $\ell$, and $k$ is the interaction strength, also referred to as the chaos parameter, that can be tuned such that there is a 
transition from regular to deterministic chaotic dynamics in the disorder free Hamiltonian, $\tau$ 
is the time period of delta kicks that is set to unity, and 
$p$  is fixed to $\pi/2$ to compare with the known results.  
Here, $\epsilon_{\ell \ell'}$ are the random part (quenched disorder) of spin interactions
taken from a normal distribution with zero mean and standard deviation
$w$, also called disorder strength. The corresponding unitary Floquet operator $U_{w}$ that evolves states infinitesimally before consecutive kicks separated by the period $\tau$, is given by
\begin{equation}
\exp\left ({-i\frac{k}{2N}\sum_{\ell< \ell'=1}^{N}(1+\epsilon_{\ell \ell'})\sigma_{\ell}^{x}\sigma_{\ell'}^{x} }\right) \exp\left ({-i\frac{\pi}{4}\sum_{\ell=1}^{N}\sigma_{\ell}^{y}}\right ),
\label{eq:U}
\end{equation}
where we have set $\hbar=1$.  A schematic diagram of the Hamiltonian studied is shown in Fig. \ref{fig_schematic}.

\begin{figure}
\centering

\includegraphics[width=0.4\linewidth]{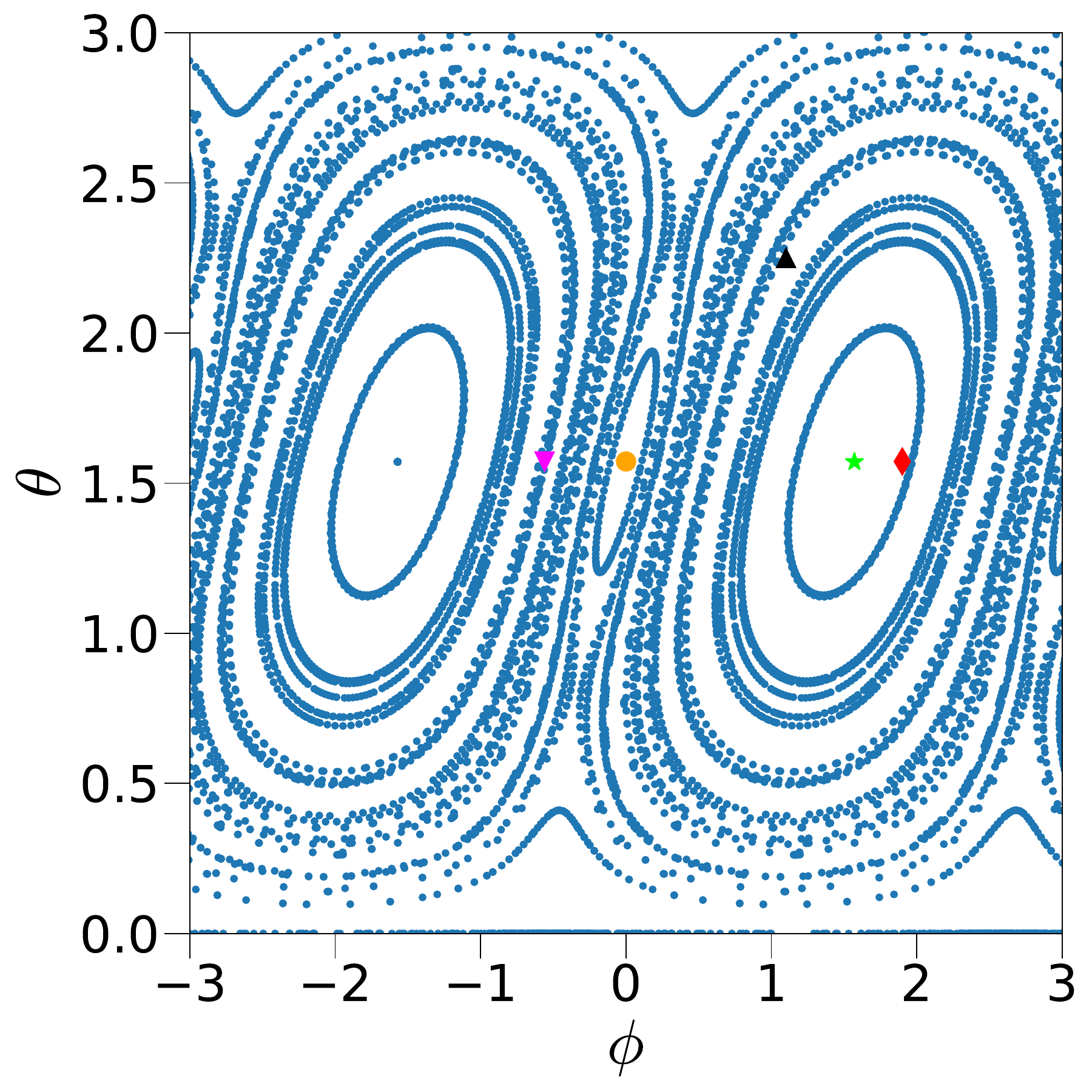}
\centering
\caption{Classical map of kicked top for $k=1$ and $p=\pi/2$. The points highlighted in various colors correspond to different initial conditions $(\theta, \phi)$ studied in this work, namely,  $\blacktriangle = $ $(2.25,1.1)$, \textcolor{orange}{$\bullet$} = $(\pi/2,0)$ ,\textcolor[HTML]{00FF00}{$\star$} =$(\pi/2, \pi/2)$ ,  \textcolor{magenta} {$\blacktriangledown$} = $(\pi/2,-0.56)$, \textcolor{red}{\ding{117}} = $(\pi/2,1.9)$. }
\label{phase_space}
\end{figure}

The disorder free version $U_0$, with all $\epsilon_{\ell \ell'}=0$,  is the quantum kicked top \cite{Haakebook, Sandrobook, haake1987classical, zyczkowski1990indicators, effect_measurement_qkt, periodicity, fox1994chaos, spinsqueezing, zou2022pseudoclassical, kickedpspin, quantummetrology, bifurcation_QKT, sieberer2019digital, papparaldi_bridging, periodic_orbits, lombardi2011entanglement, Amit_Anand_2023}
where the Hamiltonian is purely a function of the collective spin operators $J_{\alpha}=\sum_{\ell}^N \sigma_{\ell}^\alpha/2$ where $\alpha=x,~y,~z$. 
The total angular momentum $J^2=J_x^2+J_y^2+J_z^2$ defined in terms of these collective 
spins is a constant of motion in the quantum kicked top 
\cite{Haakebook, concurrence, decoherence}. Various aspects of disorder free long range interacting systems such as the one given above are discussed in Ref. \cite{defenurmp,defunu24, pappalardi2018scrambling}. The large $N$ limit of $U_0$ is studied as a symplectic classical map on the unit sphere $(X,Y,Z) \mapsto (X',Y',Z')$ where $X=J_x/j$, $Y=J_y/j$ and $Z=J_z/j$. 
This map shows regular dynamics for small values of $k$, typically $<2$,
and fully chaotic dynamics for large  values of $k$ ( $k>6$), where the Lyapunov exponent may be approximated as ($\ln k-1$) \cite{constantoudis1997lyapunov,wang2021multifractality}.
The Heisenberg evolution equations of the angular momentum operators are given by:
\begin{equation}
\left \langle J_{i} \right \rangle_{n+1}=\left \langle U_0^{\dagger}J_{i}U_0 \right \rangle_{n}.
\end{equation}
The classical equations of motion can be obtained in the limit $j\rightarrow \infty$, and are given by :
\begin{equation}
\begin{split}
X_{n+1}&=Z_{n},\\
Y_{n+1}&=Y_{n}\cos(kZ_{n})+X_{n}\sin(kZ_{n}),\\
Z_{n+1}&=-X_{n}\cos(kZ_{n})+Y_{n}\sin(kZ_{n}).
\end{split}
\label{map}
\end{equation}
 The classical phase space map for $k=1$ is shown in Fig. \ref{phase_space}, where ($\theta, \phi$) corresponds to the spherical coordinates of the total angular momentum, showing a predominantly regular behavior at this parameter value.

In the quantum case, one can intialize the system in a spin coherent state given by  \cite{glauber_spin_coherent_state, spincoherentstates,R.R_puri}
\begin{equation}
\ket{\psi_0}=\ket{\theta,\phi}=  \left (\cos\frac{\theta}{2}\ket{0}+e^{i\phi}\sin\frac{\theta}{2}\ket{1}\right)^{\otimes{N}}.
\label{eq_coherent}
\end{equation}
It corresponds to a localized state around the classical point $(X,Y,Z)=(\sin \theta \cos\phi, \,\sin \theta \sin \phi,\, \cos\theta)$.
The quantum state after $n$ Floquet periods is $\ket{\psi_n}=U_0^n \ket{\psi_0}$.
The states at all times $n$ are restricted to the 
$N+1$ dimensional PSS and conserve the total angular momentum $J^2=(N/2)(N/2+1)$.
It has also been shown that the chaotic phase can be described using a random matrix theory (RMT) in $N+1$ dimensions \cite{Haakebook}. For small values of $N (\leq 4)$, the kicked top is exactly solvable and does yield interesting insights already into the large $N$ cases \cite{pattnayak,fewbodykickedtop,otocandloschmidt}.

Introducing  disorder in the system will break the permutation symmetry of the Hamiltonian, and the state $| \psi_n \rangle$ after $n$ kicks given by $U_{w}^n \ket{\psi_0}$ will now be out of PSS
where $J^2$ is no longer a constant of motion.
If the state $\ket{\psi_n}$ in the presence of large disorder tends to a random state on the full Hilbert space, denoted as $\ket{\psi_{\text{RMT}}}$, the expectation value of $J_{z}^2$ can be estimated as:
\begin{equation}
\left \langle\ J_{z}^2 \right \rangle_{{\text{RMT}}} =\sum_{\ell=1}^{2^N}\lambda_{\ell}^2 \overline{|\left \langle \psi_{\text{RMT}}| \phi_{\ell} \right \rangle|^{2}}=N/4, 
\end{equation}
where $\ket{\phi_\ell}=\ket{j_1 \cdots j_N}$, $j_i=\pm 1$,
are eigenvectors of $J_z$ with eigenvalues $\lambda_\ell=(\sum_{i} j_{i})/2$. The ensemble average over random states $\overline{|\left \langle \psi_{\text{RMT}}| \phi_{\ell} \right \rangle|^{2}}=1/2^N$ is used above along with the fact that
$\mbox{tr} J_z^2= \sum_{\ell=1}^{2^N}\lambda_{\ell}^2= N 2^{N-2}$. Similarly  $\left \langle J_x^2 \right \rangle_{\text{RMT}}=\left \langle J_y^2 \right \rangle_{\text{RMT}}=N/4$, and hence $\left \langle J^2 \right \rangle_{\text{RMT}}=3N/4$. 
Clearly, ${\langle J^2 \rangle}$  exhibit  a change in the scaling when the dynamics shifts from the permutation symmetric subspace where $\langle J^2 \rangle \sim N^2$ to the FHS where  the state being close to random has a  $\sim N$ scaling.
It is this transition of the dynamics to the ``full" Hilbert space that we shall focus in the current work.

One can also capture this change in the dynamics by studying von Neumann entropy $S_Q$ of a subsystem 
consisting of $Q-$ spins or qubits. The von Neumann entropy is given by $S_Q=-\mbox{tr}(\rho_Q \log_2 \rho_Q)$, 
where $\rho_Q$ is the reduced density matrix of the subsystem. The expression for $S_Q$ using ensemble averaging over 
random permutation symmetric states is approximated by \cite{tripartitescrambling}
\begin{equation}
\left \langle S_{Q} \right \rangle _{PSS} \approx \log_{2}(Q+1)- \frac{2}{3}\left(\frac{Q+1}{N-Q+1}\right).
\end{equation}
We have checked numerically that  $S_Q$ even in the regular regime of small $k$  within permutation symmetric subspace is proportional to $~ \log_2 (Q+1)$. A related work calculating the von Neumann entropy in PSS is given in Ref.\cite{castro2013entanglement}. On the other hand, when the dynamics traverse the full Hilbert space, $S_Q$ is given by the Page value \cite{page1993average, tripartitescrambling, Vonneumannscaling}:
\begin{equation}
\left \langle S_{Q} \right \rangle _{2^{Q}} \approx Q- \frac{1}{\ln2}\left(\frac{2^Q}{2^{N-Q+1}}\right).
\end{equation}

Thus, we conclude that as the disorder strength $w$ is increased, various expectation values will potentially  shift from their values within the permutation symmetric subspace to the RMT
values relevant to the full Hilbert space. 
This is because increasing the strength of the disorder $w$ will not only take the system out of the permutation symmetric subspace but will also make it more chaotic since increasing $w$ is equivalent to 
increasing $k$ as can be seen from Eq. \ref{eq_spin}.
Below, we substantiate these change in scalings of $J^2$ and
$S_Q$ using numerics.

\section {Results and Discussions}
\subsection {Solvable case of $p=0$, generic initial coherent states}
It is both instructive and interesting to first examine the evolution of the model in Eq.~\ref{eq_spin} without the periodic kicks ($p=0$). This is the well-known spin-glass model, the Sherrington-Kirkpatrick model \cite{sherington_model, kirkpatrick_infinite_range, silvia_sk}, with a nonzero average interaction strength. 
The thermodynamics of this model has long been of great interest. {One can exactly solve the time evolution of disorder averaged $J^2$ denoted as   $ \langle  J^2(t)  \rangle_{w}  = \langle    J_x^{2} (t)+ J_y^{2}(t) + J_z^{2}(t) \rangle_w $, where $J^2(t)=\langle \psi(t)|J^2|\psi(t) \rangle$. Here, $|\psi(t)\rangle$ is the state at time $t$ starting from an initial state $|\theta, \phi \rangle$ in the PSS.
Since $J_x^2$ commutes with the $p=0$ Hamiltonian,
\begin{equation}
\langle  J_x^{2} (t) \rangle_w=  \frac{N}{4} + \frac{N(N-1)}{4}\cos^2{\phi}\sin^2{\theta}
\end{equation}

The calculation of $\langle   J_y^{2}(t)+J_z^{2} (t)\rangle_w$ is more involved. However, a detailed analysis (see Supplementary materials for details) leads to:
\begin{equation}
\langle  J_y^{2} (t)+ J_z^{2} (t)  \rangle_w = \frac{2N}{4}+\frac{N(N-1)}{4}(1-\cos^2 \phi \sin^2 \theta) \left [\int_{-\infty}^{\infty} \frac{1}{\sqrt{2 \pi w^{2}}}\cos\left(\frac{kt}{N}\epsilon_{ij}\right)\exp\left(-\frac{\epsilon_{ij}^2}{2w^2}\right) d\epsilon_{ij} \right]^{2(N-2)}\nonumber
\end{equation}
\begin{equation}
= \frac{2N}{4}+\frac{N(N-1)}{4} (1-\cos^2 \phi \sin^2 \theta) \exp\left[-w^2 k^2 t^2 (N-2)/N^2\right], 
\end{equation}
and hence finally
\begin{equation}
\langle  J^2 (t)  \rangle_w = \frac{3N}4 + \frac{N(N-1)}4 \left[ \cos^2 \phi \sin^2 \theta \right. + 
 \left. (1-\cos^2 \phi \sin^2 \theta) \exp\left[-w^2 k^2 t^2 (N-2)/N^2\right] \right].
\label{eq:solvable}
\end{equation}

Clearly, for a generic initial coherent state, $\langle J^2 (t) \rangle_w$ decreases exponentially with time and saturates to a disorder independent, but initial state dependent value, which is $O(N^2)$. The irreversible nature of the dynamics arises from the disorder averaging even in finite systems. When $\theta=\pi/2$ and $\phi=0$ or $\pi$, there is no decay and $\langle J^2 (t) \rangle_w=\langle J^2 (0) \rangle=j(j+1)$, where $j=N/2$. These special initial states have no evolution except for a phase as they are eigenstates of the Hamiltonian. 
Interestingly, when $\phi=\pm \pi/2$ (any value of $\theta$), 
$\langle J^2(t) \rangle_w$ saturates to the RMT value given by $3N/4$  for any non-zero value of disorder. These initial states are in the $y-z$ plane of the Bloch sphere, and are maximally {\it coherent} (in the sense of quantum information theory, as discussed below) in the preferred $x$ basis as this is the direction of the interaction. These are the only initial states for which the saturation value of $J^2$ scales linearly as $N$, rather than quadratically, and are unstable in the sense that any deviation in the $\phi$ value changes this from the RMT scaling.
We will find that in the presence of the transverse field kicks when $p \neq 0$, these behaviours are interestingly modified, while some persist.
Below, we present our results in the presence of kicks, first for a generic coherent state and then for a non-generic, initial coherent states.

\begin{figure}
\centering
   \includegraphics[width=0.45\linewidth,height=0.4\linewidth]
  {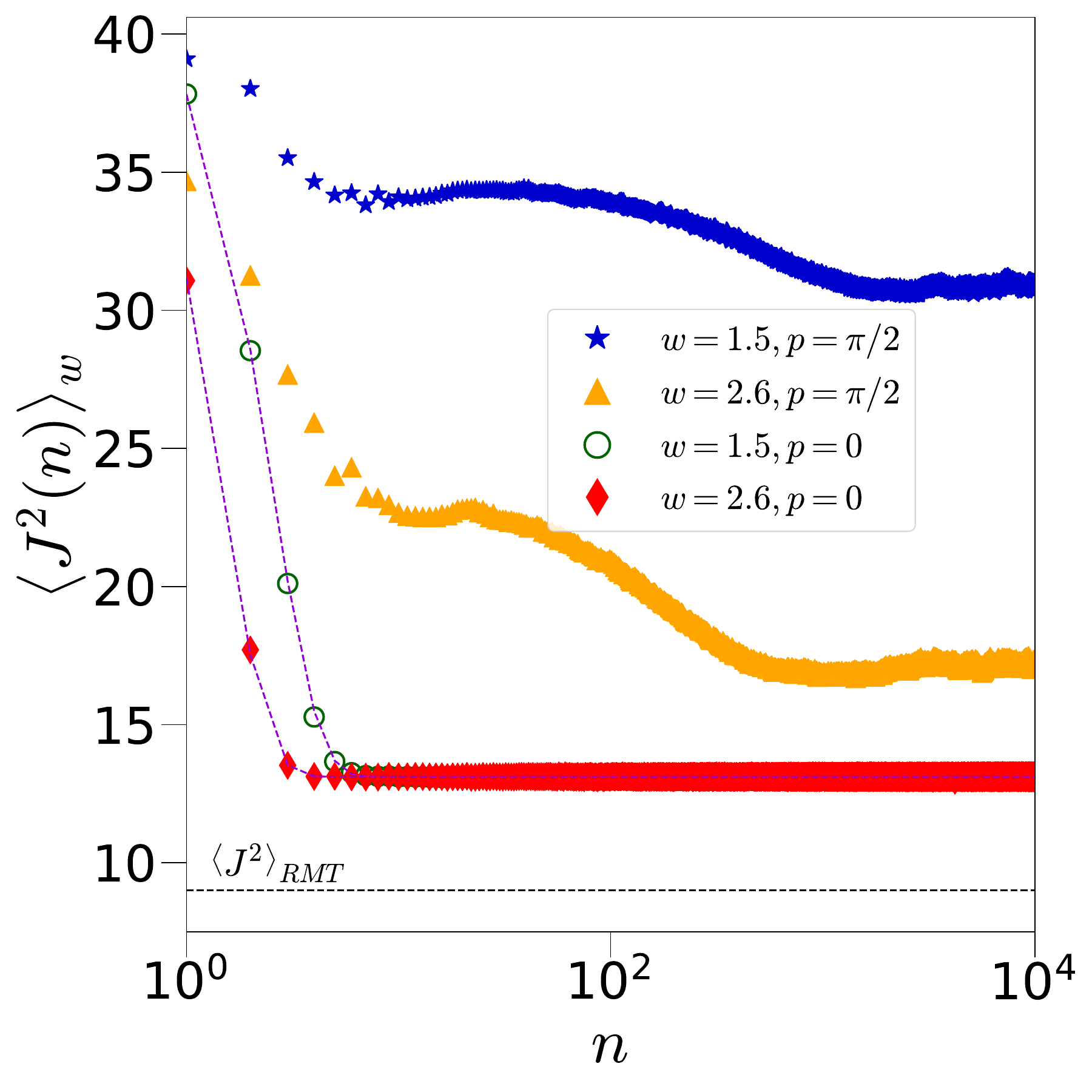} 
    \caption{ Evolution of $\left \langle J^{2}(n) \right \rangle_w$ with time $n$ for the disordered system with kicks $(p=\pi/2)$ and without kicks $(p=0)$ for two different values of disorder strength $w=1.5,~2.6$, initial state $\ket{\theta,\phi}=\ket{2.25,1.1}$, $k=1$, and $N=12$. The points correspond to numerical data and the dotted lines correspond to the analytical expression given in Eq. \ref{eq:solvable} for the unkicked Hamiltonian. The black horizontal lines is $\langle J^2 _{RMT} \rangle =3N/4=9$ which is well below the saturation value. For $p=0$, the saturation value clearly is independent of the strength of disorder, but depends on the initial state, as also predicted by Eq. \ref{eq:solvable}. On the other hand, when $p=\pi/2$, saturation value depends upon $w$ and initial state.}
    \label{comparison}
\end{figure}

Before doing that, it is noted that the kicked top unitary as defined in Eq.~(\ref{eq:U}), in the absence of disorder, will be strictly periodic in $k$, with the period $4\pi N$. Thus, the limits $k=\infty$ and $p=0$ are not equivalent, and the analytical scaling found for the latter case does not continue to hold for large $k$. In fact, we will restrict values of $k$ to be much smaller than this periodicity. Also comparing the  limiting cases for $p$ and $k$ is complicated by the Dirac delta kicks. In the usual kicked top without disorder, the relevant limit is $N \rightarrow \infty$ with a fixed $k$ and $p$.

\begin{figure}[t]
\centering
\includegraphics[width=0.4\linewidth]{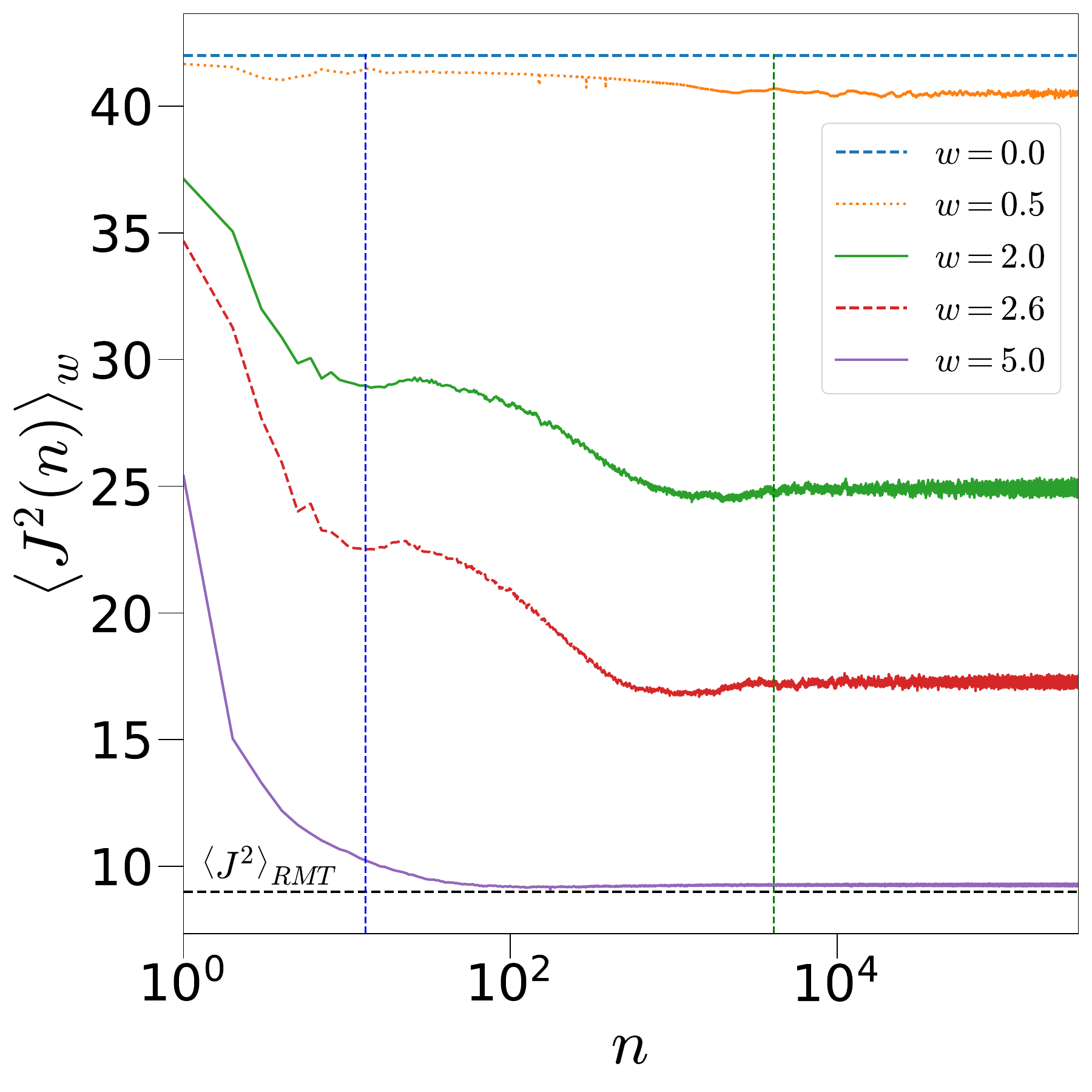}%
\includegraphics[width=0.4\linewidth]{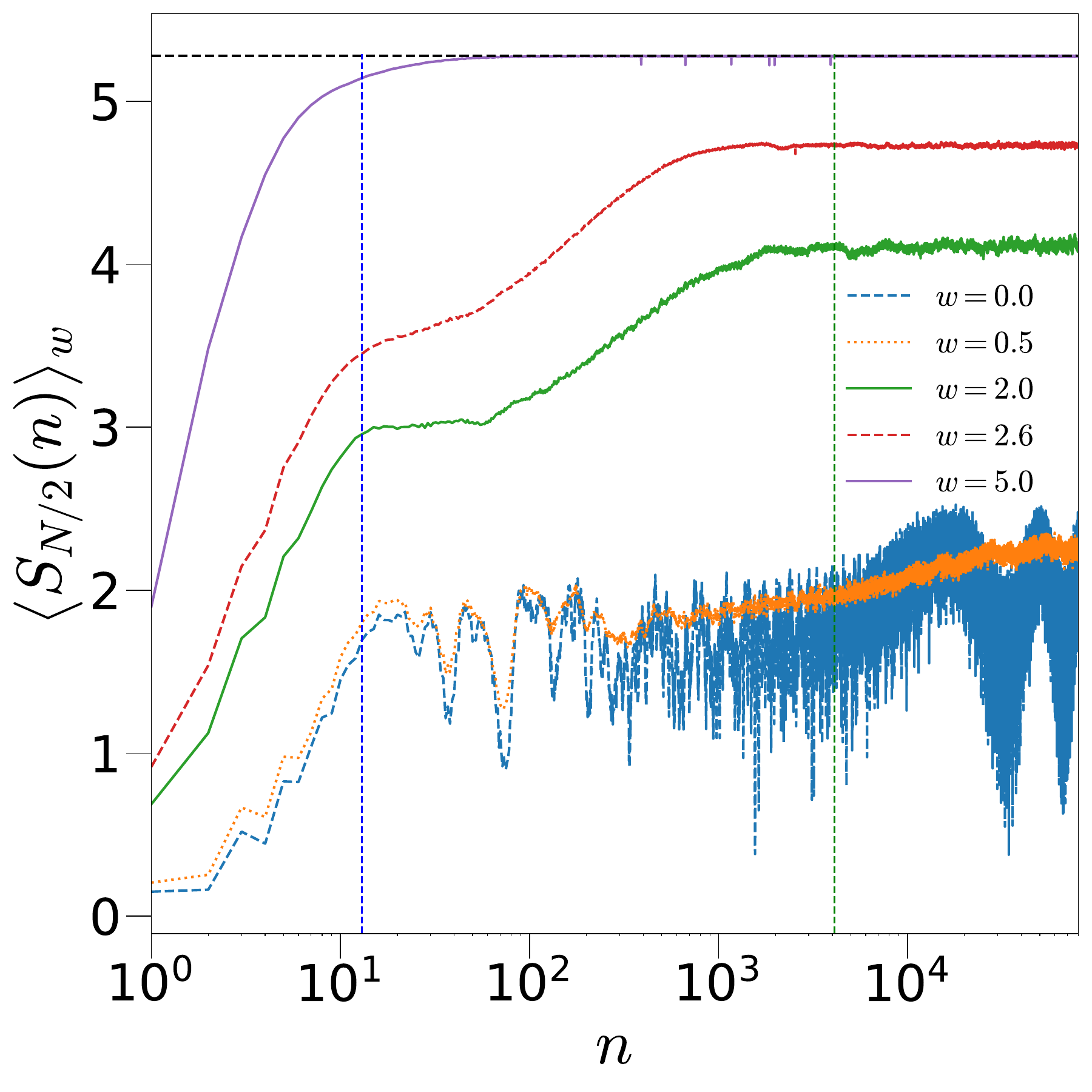}
\caption{Left: $\left \langle J^2(n) \right \rangle_w$ as a function of  number of kicks $n$ for different disorder strengths $w$ with  $k=1$, $p=\pi/2$, $N=12$ qubits, averaged over 100 disorder realizations. Right: shows the von Neumann  entanglement entropy $\langle S_{N/2} \rangle_w$ as a function of  $n$ for the same parameters. 
The two vertical lines correspond to the two Heisenberg time scales discussed in the text, $t_{PSS}^H$ (blue) and
$t_{FHS}^H$ (green). The black dashed horizontal line in both figures correspond to the RMT values in full Hilbert space. The initial state considered is $\ket{2.25,1.1}$.}

\label{withtime}
\end{figure}

\subsection {Case of $p=\pi/2$, generic initial states}
\label{sec_gen}

We present the results of numerical calculations which are performed with an  initial state $\ket{\psi_0}$ in the PSS as given in 
Eq. \ref{eq_coherent} with $\theta=2.25,~\phi=1.1$ ($\blacktriangle$ in  Fig. \ref{phase_space})
and is evolved to $\ket{\psi_n}=U_w^n \ket{\psi_0}$ using the Floquet 
operator $U_{w}$ given in Eq.~\ref{eq:U}. This initial state is ``generic" in the sense that it has no special dynamical significance in the permutation symmetric classical limit. It is still a coherent state and not generic in the sense of Haar measure.
We use the fast Walsh Hadamard transform to reach system sizes of $N=16$, and in one case to $N=18$. 
Fig. \ref{comparison} compares the dynamics of the system, both, in the absence $(p=0)$ and in the presence of kicks $(p=\pi/2)$ for two different values of disorder strength $w$. In the absence of kicks, and for the generic initial coherent state with $\cos \phi \neq 0$, the system reaches its saturation value when time $t>t^* \sim \sqrt{N}/wk$  in the large $N$ limit (see Eq. \ref{eq:solvable}).
The relevant time scale for saturation in the presence of both disorder and kicks is much more and is shown in
 Fig. \ref{withtime} using both the average total angular momentum $\langle J^2(n) \rangle_w$ and the entanglement entropy $\langle S_{N/2} (n) \rangle_w $ for $k=1$. 
 
 In contrast to the unkicked model, there is a disorder dependent steady state value eventually reached for all values of the disorder; however we find an interesting feature consisting of two regimes, first is till time $n \sim t^H_{PSS}= N+1$ corresponding to the Heisenberg time relevant to the PSS, when there is a quasi steady  state value. The other is at  $n \sim t^H_{FHS}= 2^N$, the Heisenberg time relevant to the FHS where $\langle J^2(n) \rangle_w$ and $\left \langle S_{N/2}(n) \right \rangle_w$ again starts saturating, which implies that the system has ``realized" the presence of the full Hilbert space. Notice that with large disorder strength such as $w=5$, when the full Hilbert space is being accessed, the time scale at $t^H_{FHS}$ has become irrelevant for this quantity. Instead the Heisenberg time for the PSS provides a time scale for thermalization, when $\langle J^2(n) \rangle_w$ and $\left \langle S_{N/2}(n) \right \rangle_w$ saturates close to the RMT value corresponding to FHS. This is to be expected as the Heisenberg time for the PSS is also the Ehrenfest or log-time for the dynamics in the FHS. For intermediate values of the disorder strength $w$, the nonequilibrium state's $J^2$ as well as the entanglement saturates to non-thermal (non-RMT) values, signifying some kind of localization. It would be interesting to compare these with the well-studied case of MBL \cite{MBL_pal, oganesyan_pal, MBL_Luitz}.
\begin{figure}[t] 
\centering
    \includegraphics[width=0.4\linewidth]{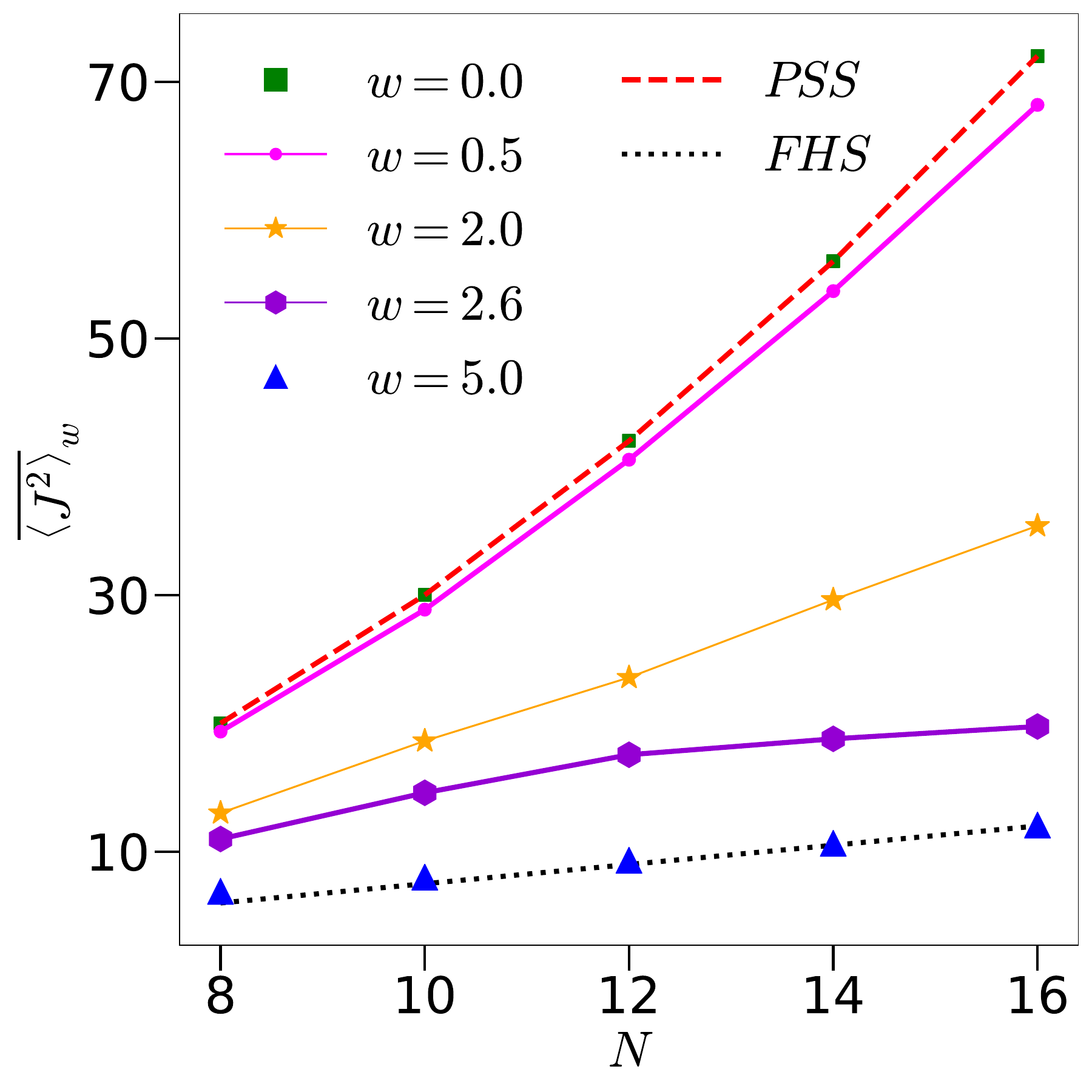} %
    \includegraphics[width=0.4\linewidth]{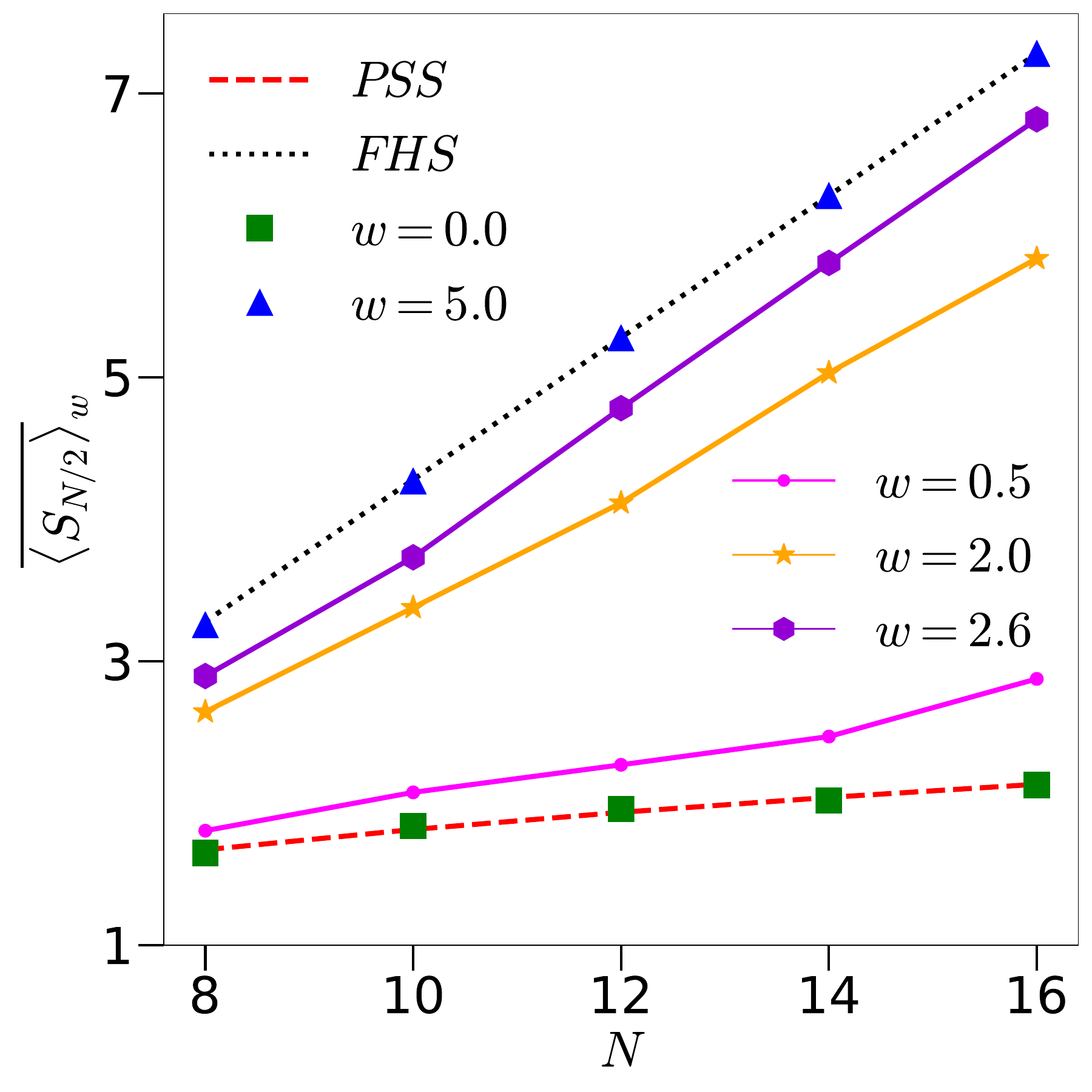} 
    \caption{Left: Change in scaling of $\overline{\langle J^2 \rangle}_w$  from $N^{2}/4+N/2$ (red dashed line) to $3N/4$ (black dotted line) as disorder strength $w$ is increased with $k$ set to 1 and $p=\pi/2$. Right: Same for $\overline {\langle S_{N/2} \rangle}_w$ as a function of $N$ at $k=1$ where the red dashed line corresponds to fitted line $0.41+0.54 \log_2 (N/2+1)$ and the black dotted line is the RMT expression given by $N/2- 1/2 \ln 2$. The initial state is $\ket{2.25,1.1}$.}
    \label{JvsN}  
\end{figure}

To quantify the saturation value as a function of the disorder and system size, we now study the behavior of long time averaged values (averaged from $n=10^5$ to $n=3 \times 10^5$ for 100 disorder realizations) of $J^2$ and $S_{N/2}$, denoted
as $\overline {\langle J^2\rangle}_w$ and $\overline {\langle S_{N/2} \rangle}_w$, respectively.
From the discussions above, we expect that as  $w  \to 0$ or small when the dynamics is predominantly within the permutation symmetric subspace, the scaling of long time limit of various quantities will follow the PSS scaling, $i.e.,$
$\overline {\langle J^2\rangle}_w\sim N^2$ and 
$\overline {\langle S_{N/2} \rangle}_w  \sim  \log_{2}(N/2+1)$.  In the other extreme limit, when  $ w \to \infty$ and the dynamics 
is described by RMT in the full $2^N-$dimensions, 
$\overline {\langle J^2\rangle}_w=3N/4$ and 
$\overline {\langle S_{N/2} \rangle }_w\sim N/2$. Fig. \ref{JvsN} highlights the above discussed change in scaling when $w$ is increased.
Note that for the solvable case in the absence of kicks, while Eq.~\ref{eq:solvable}, indicates an earlier saturation time scale $\sim \sqrt{N}$ for generic initial coherent states, the saturation value itself is still similar to the PSS case with $\langle J^2 \rangle \sim N^2$.

It is tempting to associate this shift in the scaling of $\overline{\langle J^2 \rangle}_w$ and $\overline {\langle S_{N/2} \rangle}_w$
with a second order phase transition.
Therefore, we now investigate the possible existence of a 
critical $w_{c}$ below which the dynamics is predominantly 
within the PSS and above which the 
dynamics captures the properties of the FHS
described by RMT, and obtain relevant critical exponents of the associated phase transition, if any. We propose a finite size scaling of the form \cite{botetpfeuty82,botet83, curveJ2similar, Vonneumannscaling, dynamicalcriticalscaling,cardy1988finite}
\begin{equation}
{\overline{\langle J^2 \rangle}_w}=N^{\zeta/\nu}F((w-w_c)N^{1/\nu}),
\label{eq_scaling}
\end{equation}
where $\nu$ is the correlation length exponent. Fig. \ref{fig_crossing} (a) shows the crossing of the data
around $w_c=2.11$ when $k=1$ whereas the collapse of the data with $\nu=0.50$ and $\zeta=0.57$ 
is shown in Fig. \ref{fig_crossing} (b). In order to confirm this phase transition further, we also study the variance of $J^2$, var$(J^2)$=$\langle(\overline{J^2}-{\langle \overline{J^2} \rangle})^2 \rangle$ where $\overline{J^2}$ is the time averaged value for each disorder. This quantity also shows a peak around $w_c=2.0$ consistent with finite size scaling results, see Fig. \ref{fig_crossing}(c) \cite{khemani2017critical,abanin2021distinguishing,variance_Sn/2}.
Next, we repeated the calculations for other values of $k$, the results of which are shown in Fig.~\ref{diff_k}.

 Table \ref{table1} lists the critical disorder strength $w_c$ along with the critical exponents $\nu$  and $\zeta$ for different values of the interaction strength or chaos parameter $k$. Intuitively, the chaotic dynamics at larger $k$ values will require smaller disorder to take the dynamics to the full Hilbert space, which is also what we observe. Interestingly, we also find a weak parameter dependence on the critical exponent $\nu$ for this phase transition. The exponent $\nu$ lies within a range of 0.6 to 0.3 for the parameters studied. Such a variation seems to be a characteristic of a finite size disordered system such as the well studied many-body-localization \cite{PRLwang, MBL_Luitz}. We also find that in most cases, $\nu \sim \zeta$.
The quality of the data collapse for different values of $k$ confirm the presence of a second order phase transition.  At the same time, we do observe that identifying correct $w_c$ values for larger $k$ is difficult since $w_c$ then is very small so that a single crossing point for all system sizes is difficult to obtain, and hence we restrict our studies for $k$ values upto $2$.
Numerics show a decrease in the value of the critical strength as $k$ is increased, indicating zero critical strength for large value of $k$.
On the other hand when $k \rightarrow 0$, the dynamics can get highly non-ergodic and these parameters need more careful examinations. 

\begin{table}[ht]

  \begin{center}
    
     \begin{tabular}{|c|c|c|c|}
      \hline
      $k$ & \textbf{$w_c$} &  \textbf{$\nu$} & \textbf{$\zeta$}\\
    \hline
      $0.5$ & $3.16 \pm 0.04$ &  $0.58 \pm 0.03$ & $0.62 \pm 0.05$ \\ 
      \hline
      $1.0$ & $2.11 \pm 0.06$ &  $0.52 \pm 0.06$ & $0.57 \pm 0.09$\\
          \hline
      $2.0$ & $0.47 \pm 0.02$ &  $0.37 \pm 0.06$ & $0.40 \pm 0.06$\\
      \hline
      \end{tabular}
  \end{center}
    
  \caption{Critical disorder strength \textcolor{blue}{$w_c$} and exponents $\nu$, $\zeta$ for different chaos parameter, $k$. We have used $\overline{\langle J^2 \rangle}_w$  with $N=12,14$ and $16$ to identify the critical points with time averaging ranging from $n=10^5$ to
    $3 \times 10^5$ for 100 different realizations. The initial state is fixed to $|\theta, \phi \rangle= |2.25, 1.1\rangle$ .
}
    \label{table1}
        
\end{table}

\begin{figure}[H]
\centering
\includegraphics[width=0.3\linewidth]{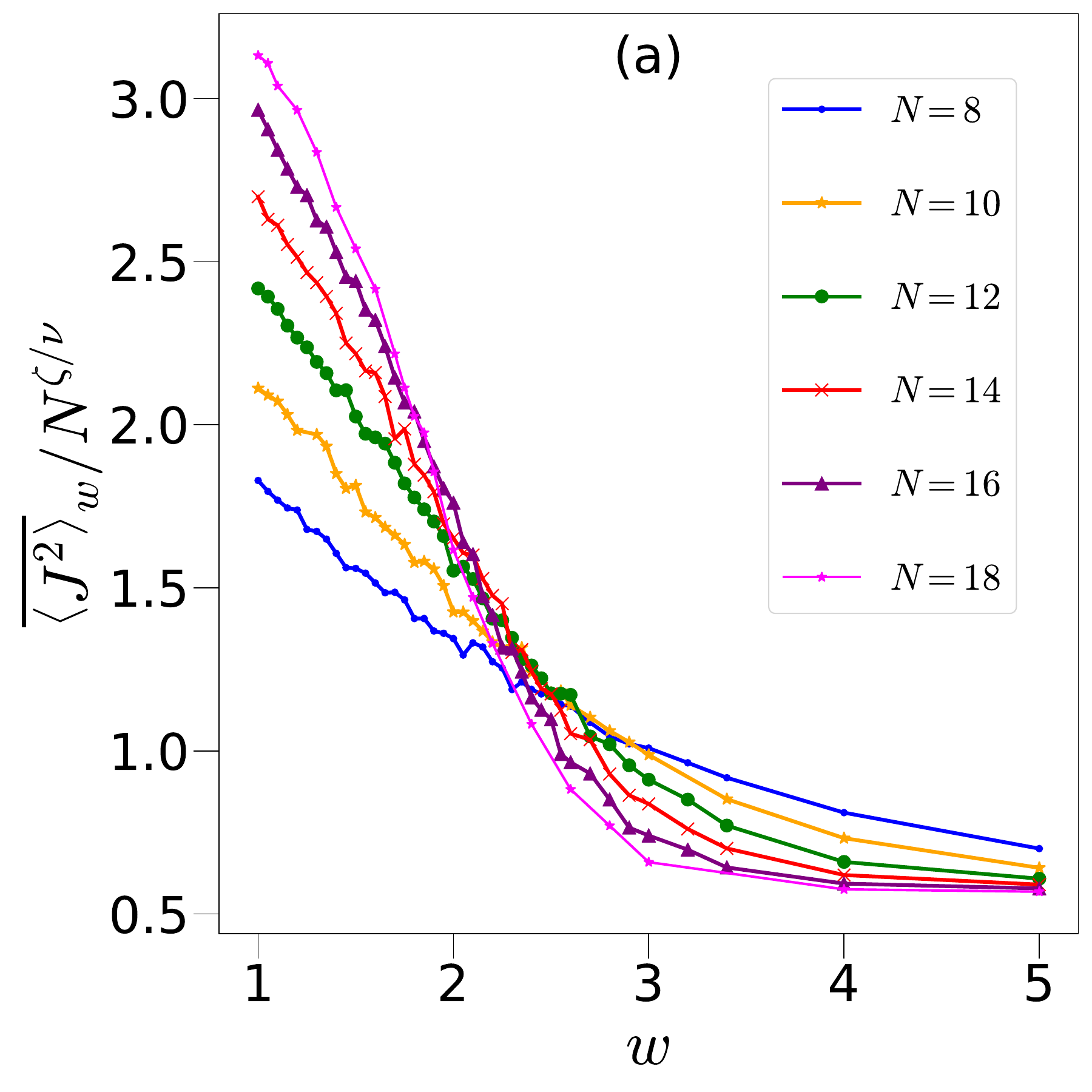}
\includegraphics[width=0.3\linewidth]{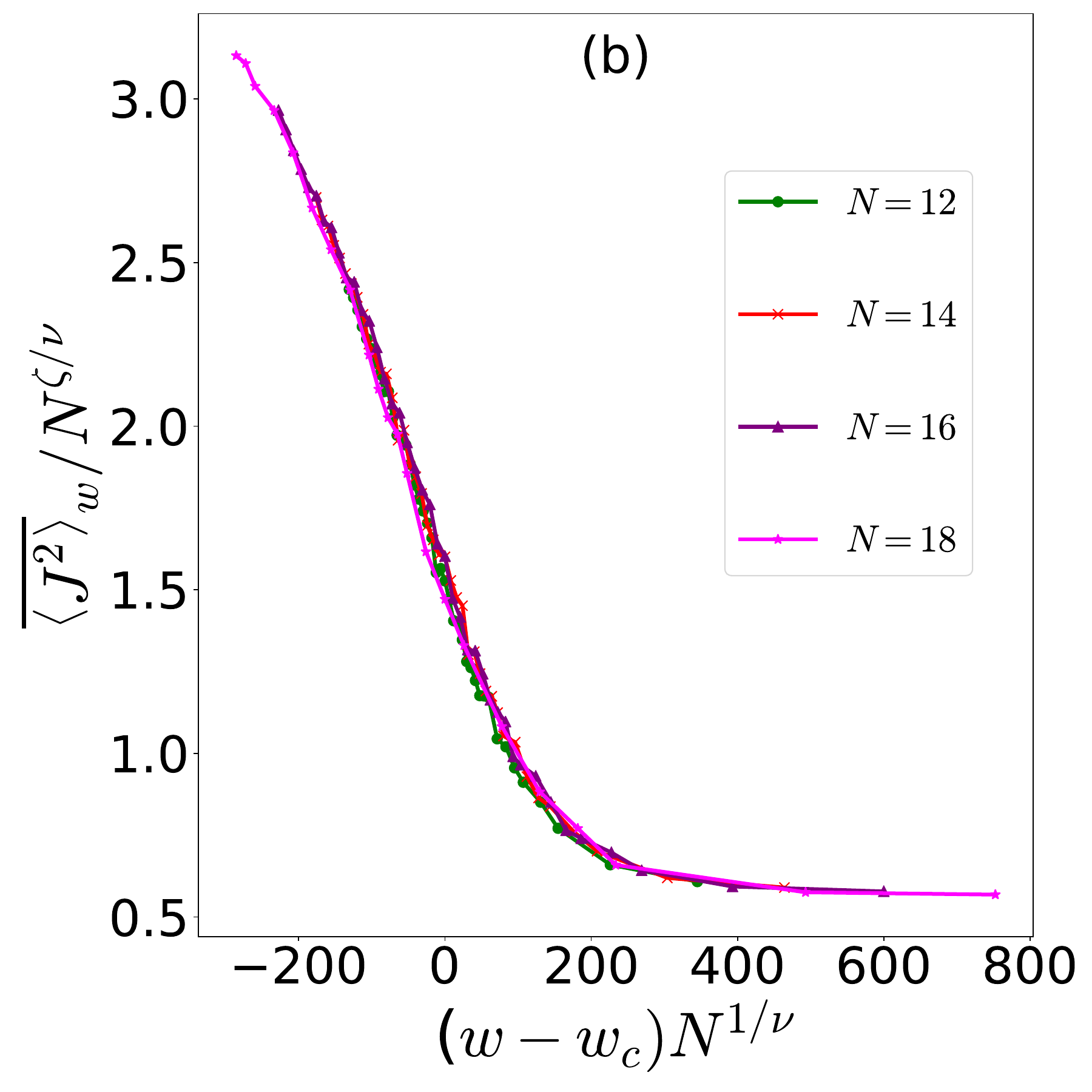}
\includegraphics[width=0.3\linewidth]{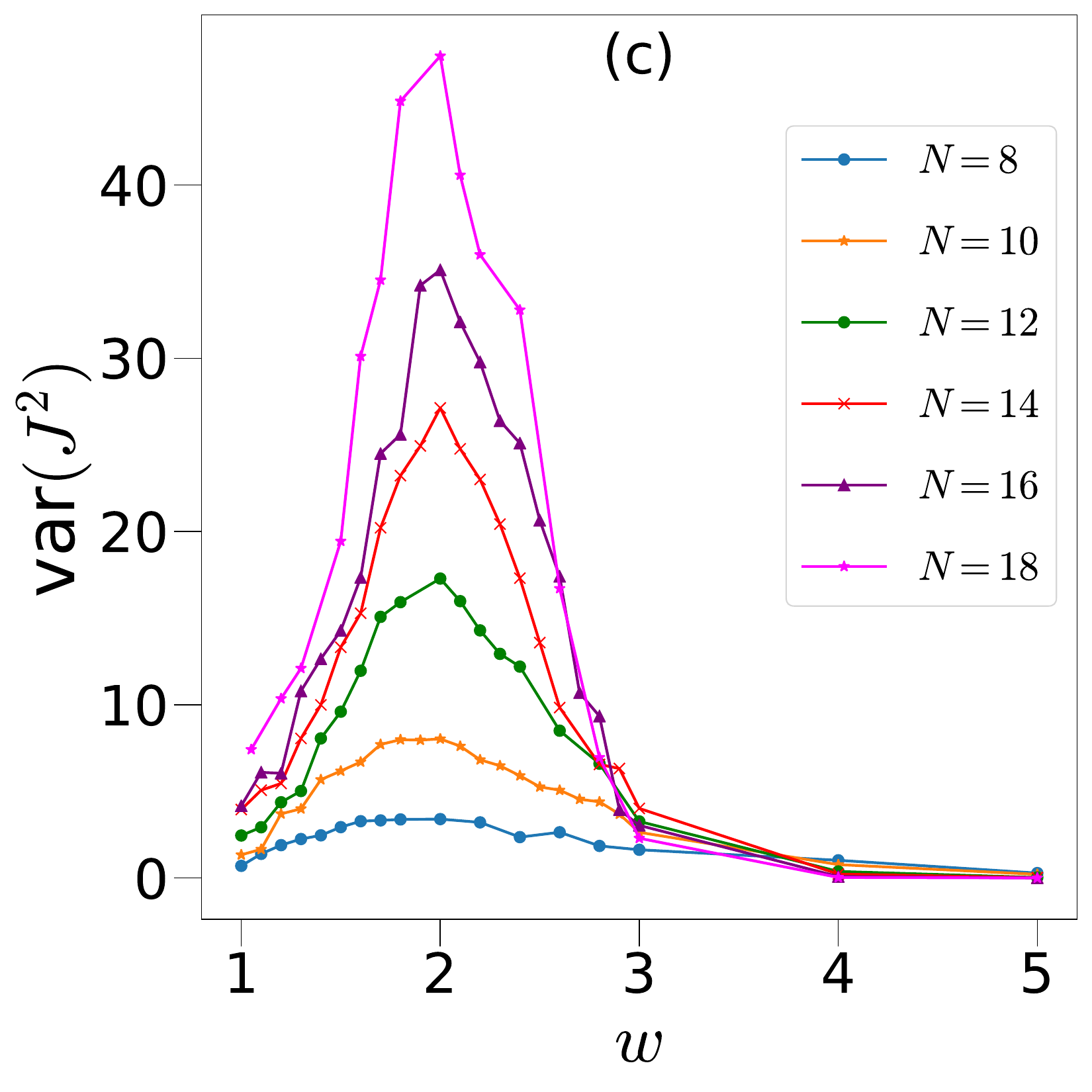}
\caption{(a): $\overline{\langle J^{2} \rangle}_w/N^{\zeta/\nu}$ (where $\zeta$ and $\nu$ are the critical exponents) as a function of disorder strength $w$ at $k=1, p=\pi/2$ and initial state $\ket{2.25,1.1}$ for different system sizes. These curves cross each other at $w=w_c \approx 2.11$. (b): The collapse of the data with $\zeta=0.57$ and $\nu=0.52$ consistent with the scaling proposed in Eq. \ref{eq_scaling}. (c): var$({J^{2}})$ with respect to disorder strength $w$ for $k=1$ that shows a peak around $2.0$. }
\label{fig_crossing}
\end{figure}
\begin{figure}[h] 
\centering
    \includegraphics[width=0.4\linewidth]{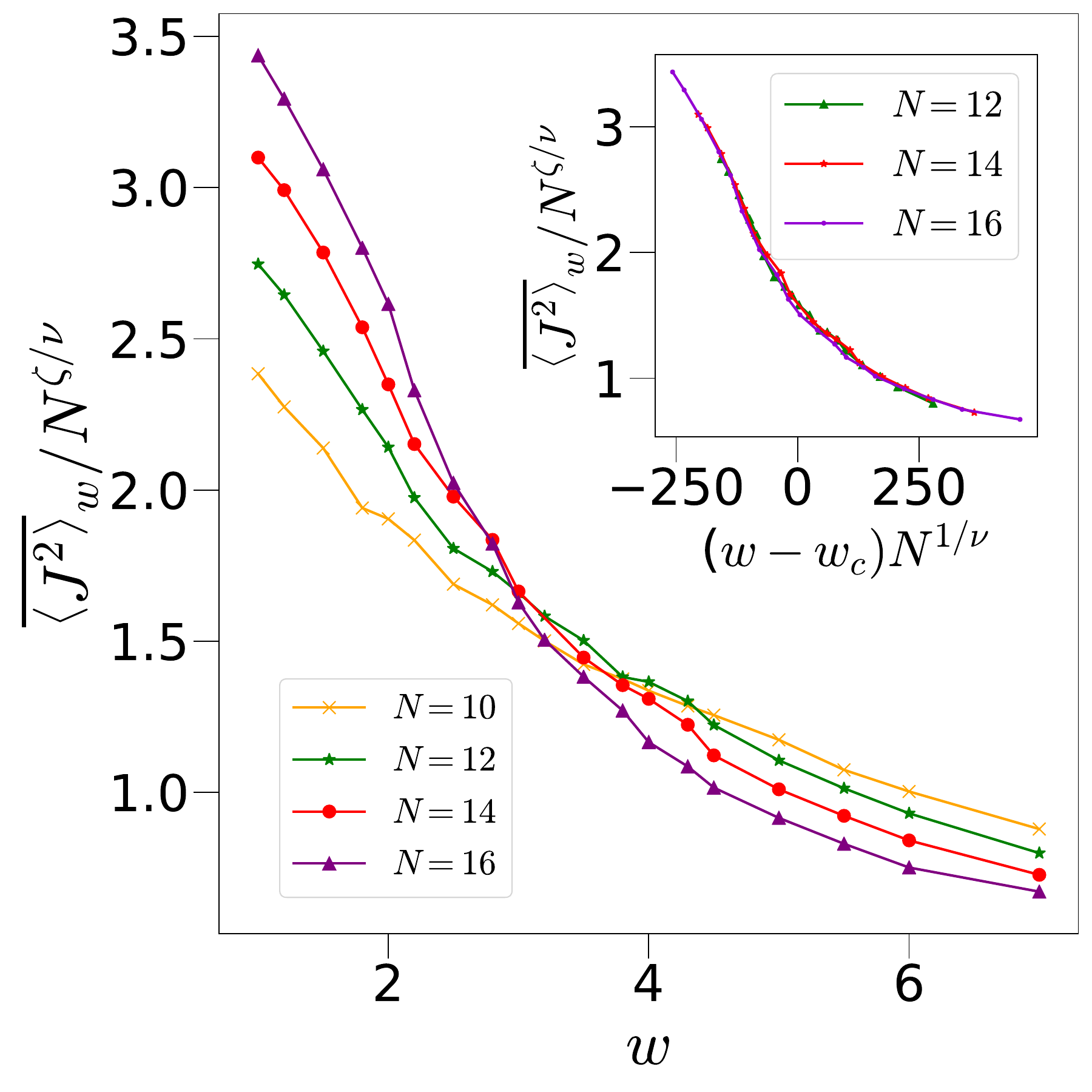} %
    \includegraphics[width=0.4\linewidth]{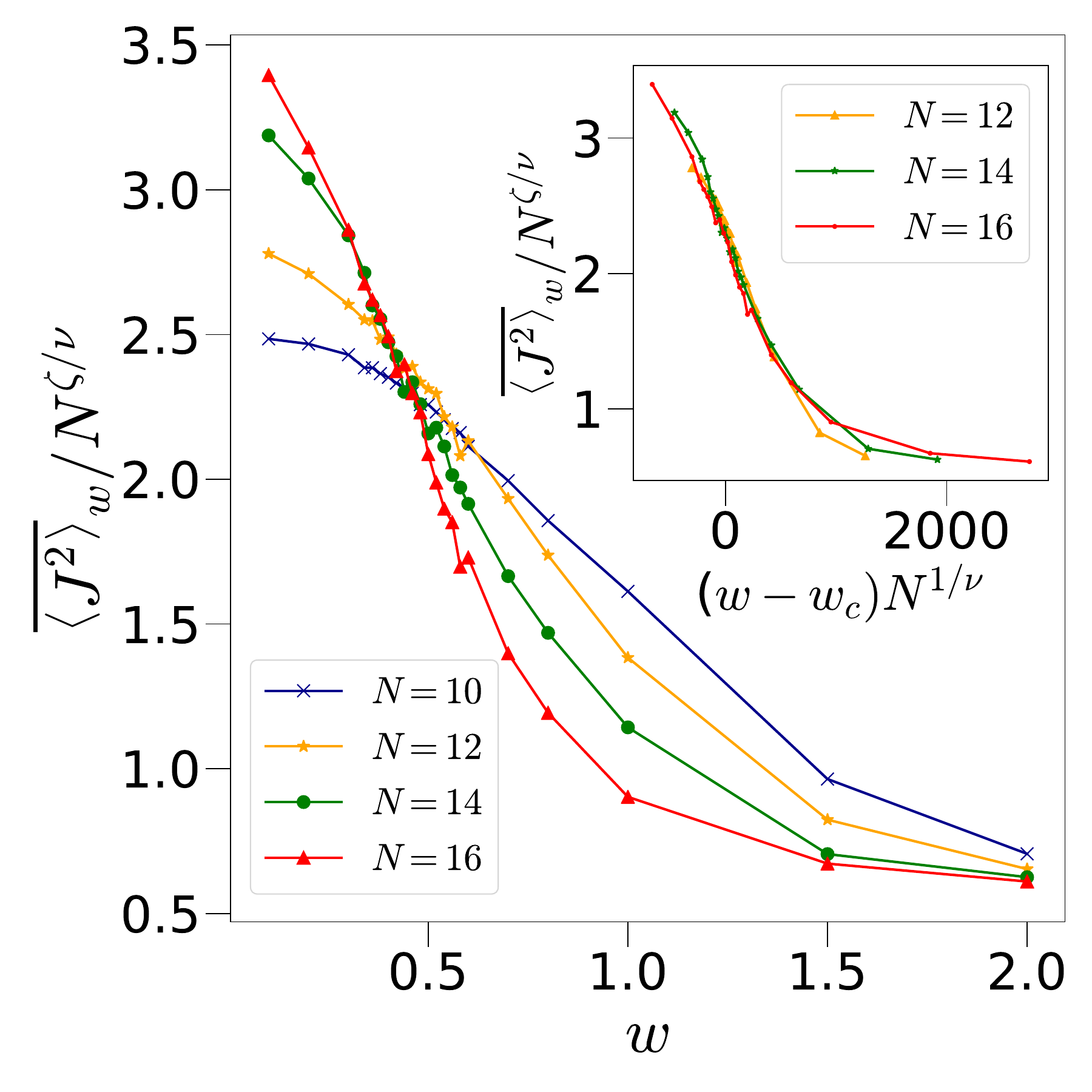} 
    \caption{(Left): The main figure shows the data of $\overline{\langle J^{2} \rangle}_w/N^{\zeta/\nu}$ with respect to $w$ corresponding to different system sizes for $k=0.5, p=\pi/2$ and initial state $\ket{\theta,\phi}=\ket{2.25,1.1}$ crossing each other at $w=w_c \sim 3.16$ whereas the inset shows the collapse of the data with $\nu=0.58$ and $\zeta=0.62$. (Right): Same as above but for $k=2$ with $w=w_c \sim 0.47$, $\nu=0.37$ and $\zeta=0.40$.}
  \label{diff_k}
\end{figure}

We have further confirmed this phase transition using entanglement entropy $\overline {\langle S_{N/2} \rangle}_w$ with a similar finite size scaling as given in Eq.\ref{eq_scaling} and a different exponent $\zeta^{'}$. The analysis gives $\nu=0.60$, $\zeta^{'}=0.68$ and $w_c=1.93$ for $k=1$, as shown in Fig. \ref{entropy}.
We do acknowledge the fact that these numerics are limited by finite size systems and may not fully capture the behavior in the thermodynamic limit. This may also be the reason for slightly different values of $w_c$ obtained using $\overline {\langle S_{N/2} \rangle}_w$ and $\overline{\langle J^2 \rangle}_w$.

\begin{figure}[h]
\centering
\includegraphics[width=0.5\linewidth,height=0.4\linewidth]{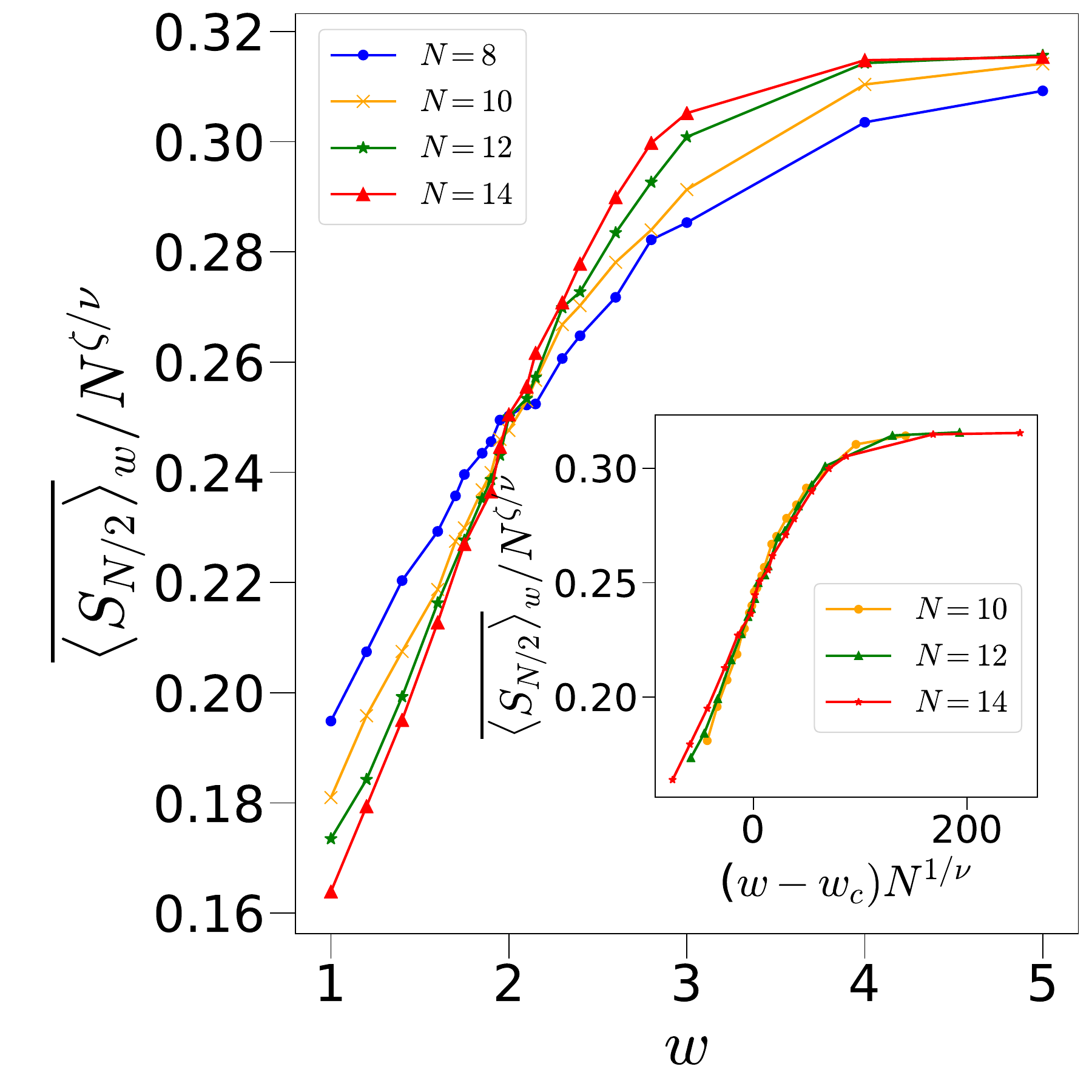}
\caption{(a) Plot of $\overline {\langle S_{N/2} \rangle }_w/N^{\zeta'/\nu}$ vs $w$ for $k=1, p=\pi/2$ and initial state $\ket{2.25,1.1}$ corresponding to different system sizes cross each other at $w=w_c\sim 1.93 \pm 0.06$. Inset shows the collapse of the data with $\nu=0.60\pm0.06$ and $\zeta'=0.68 \pm 0.07$.}
\label{entropy}
\end{figure}

\subsection {Case of $p=\pi/2$, special initial states}
The above calculations were performed for a generic initial state specified by a $(\theta, \phi)$ which has no special significance. It is interesting to study the effect of changing this initial state on the nonequilibrium dynamics and on the critical behavior. For example, in the solvable case {$p=0$}, we found that for all initial states in the $y-z$ plane of the Bloch sphere, $\langle J^2(n) \rangle_w$ eventually saturates to the RMT value, independent of the strength of the disorder. Figure \ref{diffic} shows the scaling of the saturation value $\langle J^2 \rangle$ for four different initial coherent states (see Fig. \ref{phase_space} for their location in phase space), including $(\pi/2,\pi/2)$, $(\pi/2,0)$, and two others $(\theta,\phi)=(\pi/2, 1.9)$, $(\theta,\phi)= (\pi/2, -0.56)$, which are somewhat close to $(\pi/2,\pi/2)$ and $(\pi/2,0)$, respectively.
{We observe a shift in $w_c$ for certain initial states as compared to the one reported in table \ref{table1}}.

\begin{figure}[ht] 
  \label{ fig7} 
  \begin{minipage}[b]{0.5\linewidth}
    \includegraphics[width=0.9\linewidth, height=0.8\linewidth]{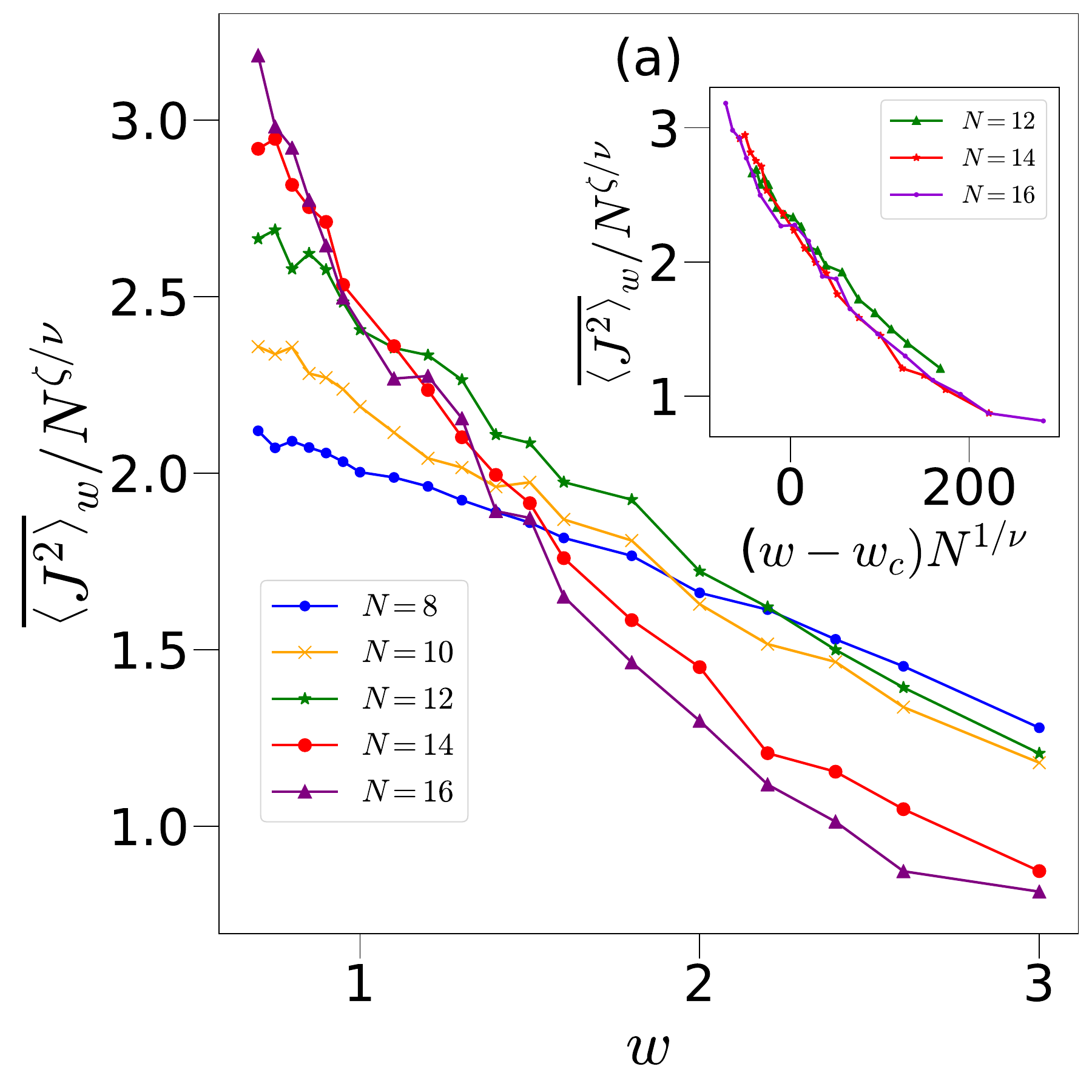} 
  \end{minipage}%
  \begin{minipage}[b]{0.5\linewidth}
    \includegraphics[width=0.9\linewidth, height=0.8\linewidth]{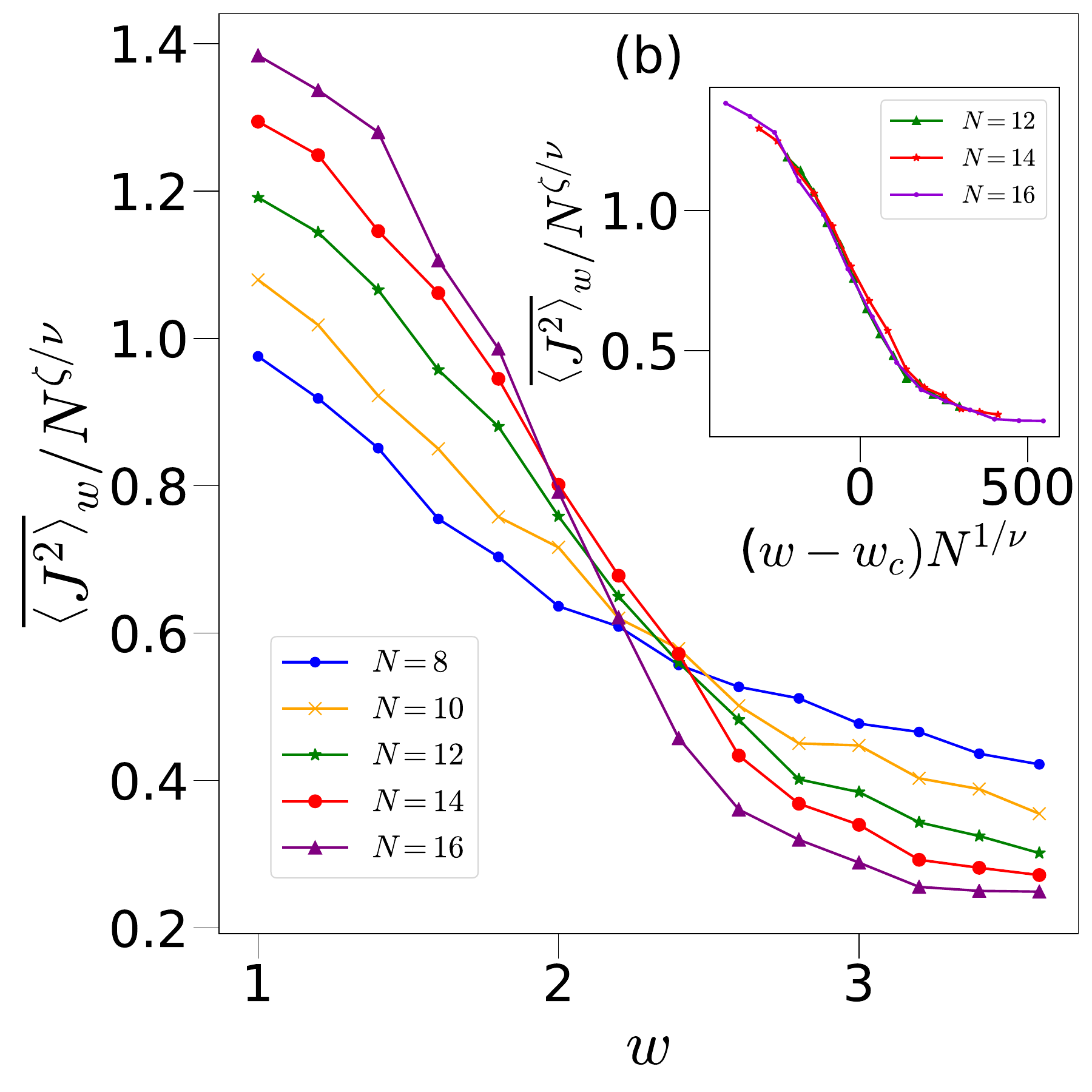} 
  \end{minipage} 
   \begin{minipage}[b]{0.5\linewidth}
    \includegraphics[width=0.9\linewidth, height=0.8\linewidth]{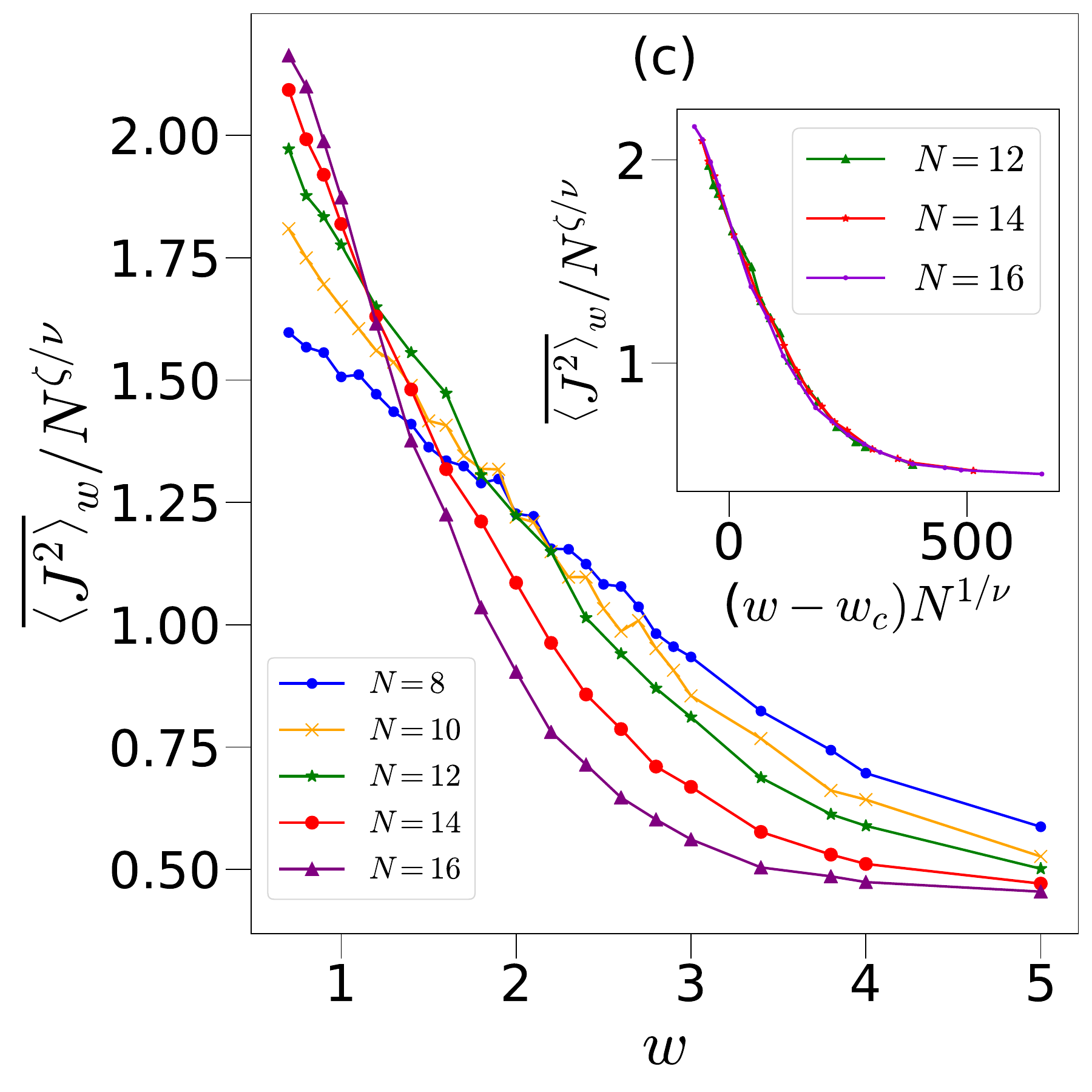} 
  \end{minipage}%
   \begin{minipage}[b]{0.5\linewidth}
    \includegraphics[width=0.9\linewidth, height=0.8\linewidth]{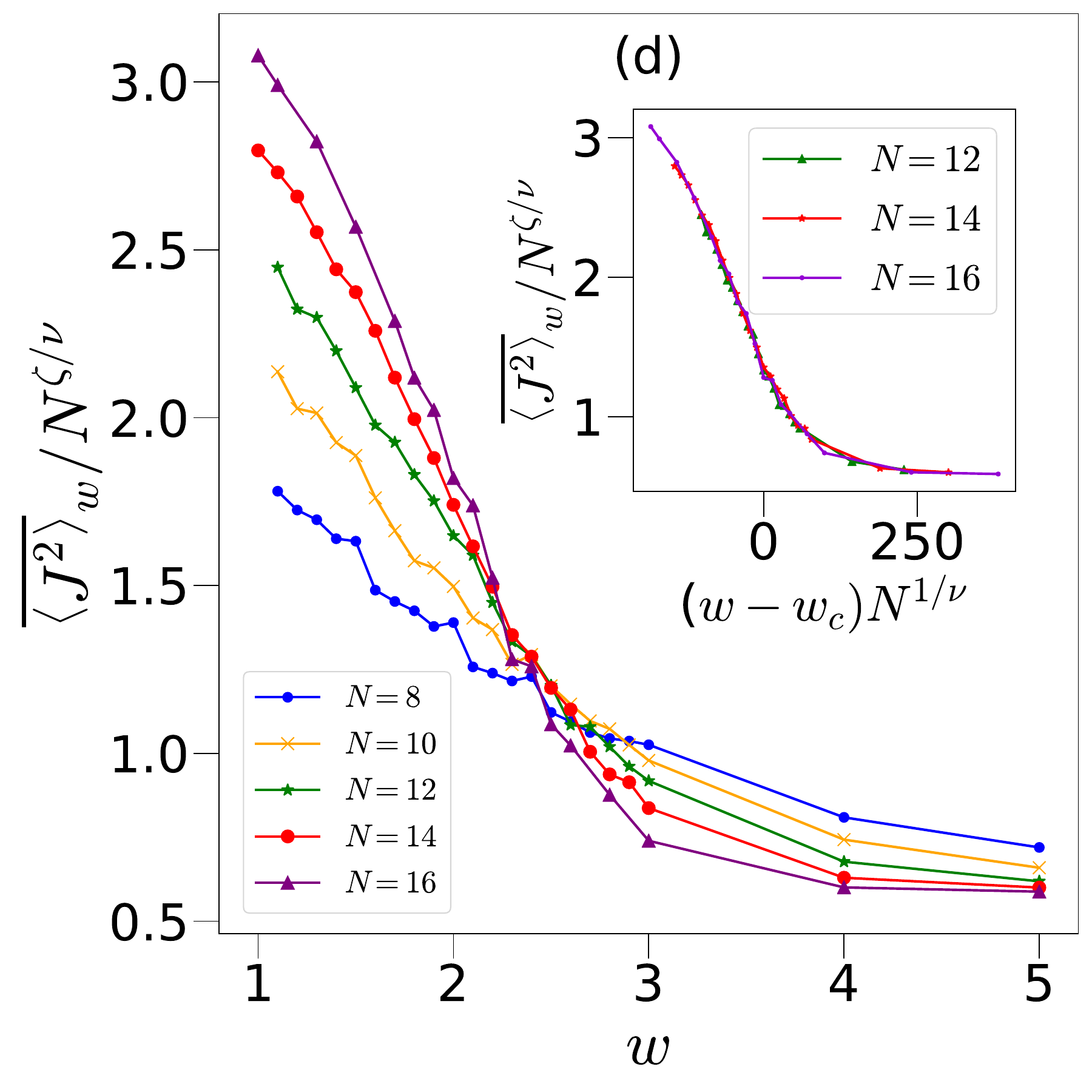} 
  \end{minipage} 

  \caption{Finite size scaling for $k=1, p=\pi/2$ and different initial conditions $|\theta, \phi \rangle$: (a) $|\pi/2, \pi/2\rangle$ shows  $w_c=1.2$, $\nu=0.55$ and $\zeta=0.58$, (b) $|\pi/2,0\rangle$ with  $w_c=2.1$, $\nu=0.47$ and $\zeta=0.66$, (c) $|\pi/2,1.9\rangle$, with  $w_c=1.10\pm0.09$, $\nu=0.54 \pm 0.01$ and $\zeta=0.64 \pm 0.06$. This initial condition is closer to $(\pi/2,\pi/2)$, and hence has a smaller $w_c$, (d) $|\pi/2,-0.56\rangle$ with $w_c=2.31\pm0.05$,  $\nu=0.56\pm 0.03$ and $\zeta=0.61 \pm 0.02$ is similar to Fig. \ref{fig_crossing}, since both the initial states are closer to $(\pi/2,0)$.}

  \label{diffic}
  \end{figure}

To understand this initial state dependence behavior better, at least qualitatively, let us start from an initial state given by $(\theta, \phi)=(\pi/2,\pi/2)$, which is an eigenstate of $\sigma_y$, so that the many body state from Eq.~\ref{eq_coherent} is fully delocalized along the $x$-direction, in which the Hamiltonian has both the interaction and the disorder. In other words the quantum coherence, as measured by the off-diagonal matrix elements of the initial state is the largest possible in the $\sigma^x$ basis. Several recent studies have shown the importance of initial state coherence as a resource for the process of thermalization \cite{Styliaris2019,Jethin2023,Mitra2023}. Thus these initial states with large coherence tend to delocalize preferentially. In the absence of disorder, the classical limit corresponding to these initial states, for $k=1$, are stable fixed points, and it is interesting that the quantum effects of disorder tend to preferentially {\it delocalize} these states. This is consistent with a lower value of the critical strength $w_c$ of the disorder. However, the crossing points in Fig.~\ref{diffic} are drifting most significantly in this case towards smaller values of $w_c$ as $N$ increases, and the scaling does not seem to work well. But with the system sizes we are able to go to, $w_c$ is not larger than $\approx 1.2$. {Infact, our heuristic argument given in the next section discusses the peculiarity of this particular initial condition which may be the reason for not getting a good value of critical strength of disorder.

On the other hand when $(\theta,\phi)=(\pi/2,0)$, the initial state is an eigenstate of $\sigma_x$, with no off-diagonal matrix elements in $\sigma^x$ basis, or zero quantum coherence. In this case, stronger disorder along the $x$ interaction direction is required to equilibriate in the full Hilbert space.  The scaling behavior seems to be more convincing as well, and the critical disorder strength $w_c$ is not larger than $\approx 2.1$. Classically this corresponds still to a stable point, however, there is a small resonance island with nearby unstable points. Thus, the coherence of the initial states seems to play a crucial role when the dynamics is regular. 
We have numerically verified this argument using other values of $(\theta,\phi)$ as well. The initial states with $(\theta,\phi)$  closer to the $(\pi/2,0)$ has a larger $w_c$ as compared to the ones closer to $(\pi/2, \pi/2)$, but with similar critical exponents, as shown in Fig.~\ref{diffic} c, d. Both these figures show relatively good crossing and collapse for $N=12, 14, 16$. 

In summary, we performed a detailed analysis for $k=1$ with different initial conditions and obtain the following critical disorder strength and the exponents by the scaling of $J^2$ and $S_{N/2}$  as shown in Table \ref{table2}:\\
\begin{table}[ht]

  \begin{center}
    
     \begin{tabular}{|c|c|c|c|}
      \hline
      $Quantity $ & \textbf{$w_c$} &  \textbf{$\nu$} & \textbf{$\zeta$}\\
    \hline
      $J^2$ $(\theta=2.25$, $\phi=1.1)$ & $2.11 \pm 0.06$ &  $0.52 \pm 0.06$ & $0.57 \pm 0.09$ \\ 
      \hline
      $J^2 (\theta=\pi/2,\phi=-0.56) $ & $2.31 \pm 0.05$ &  $0.56 \pm 0.03$ & $0.61 \pm 0.02$\\
          \hline
      $J^2 (\theta=\pi/2, \phi=1.9)$ & $1.1 \pm 0.09$ &  $0.54 \pm 0.01$ & $0.64 \pm 0.06$\\
      \hline
      $S_{N/2} (\theta=2.25, \phi=1.1)$ & $1.93 \pm 0.06$&  $0.60 \pm 0.06$ & $0.68 \pm 0.07 $\\
      \hline
     
    \end{tabular}
  \end{center}
  \caption{Critical disorder strength $w_c$ and exponents $\nu$, $\zeta$ for different initial conditions ($\theta,\phi$) for the quantities, square of total angular momentum operator $\left \langle J^2 \right \rangle$ and von Neumann entropy $S_{N/2}$ at $k=1$, $p=\pi/2$.}

    \label{table2}
        
\end{table}

 Clearly, the values of exponent $\nu$ are close to each other within the error bar, even for different initial conditions considered. The critical disorder strengths are also similar. 
 The reason for smaller value of $w_c$ for one of the initial conditions has been explained before.

\subsection{Discussions}
Given the small system sizes that we rely on in the finite size scaling analysis, a heuristic understanding may help. First, we notice that all the initial states that we take are in the permutation symmetric subspace and  the expectation value of $J^2$ scales as $N^2$, which remains exactly true at all times in the absence of disorder. 
When there is no rotation ($p=0$), we have seen that the disorder decreases the $J^2$ expectation value with time, but for almost all initial conditions the scaling remains $N^2$ in the large time limit. The exceptions are initial states lying on the great circle, $\phi=\pi/2$. Turning on the rotation about the $y$ axis in presence of small disorder, a generic permutation symmetric initial state evolves essentially approximately to other symmetric states, and the long-time expectation value of $J^2$ continues to scale as $N^2$. This is due to the fact that the rotation is itself the same for all qubits involved, in other words it commutes with $J^2$.} 

{To make this more quantitative, notice that the Floquet operator in Eq.~(\ref{eq:U}) can be written as $U_w=U_{\epsilon} U_0$, where $U_0$ is the kicked top without disorder and hence $[U_0,J^2]=0$. The  disordering part is $U_{\epsilon}$ which corresponds to a torsion about the $x-$ axis that affects individual qubits differently. If $w \ll 1$, $U_w^n \approx U_{\epsilon}^n U_0^n$. Hence, if $\ket{\psi_0}$ is a permutation symmetric initial state, $U_0^n \ket{\psi_0}$ remains in this subspace. From the analytic analysis with $p=0$ we have seen that such states which will not generally be in the exceptional great circle or even be a simple coherent state, when acted upon by $U_{\epsilon}^n$ will continue to have an expectation value of $J^2$ scaling as $N^2$. Thus, it is plausible that for small disorder values, the scaling of $J^2$ in the large time limit remains $N^2$ even when $p=\pi/2$, pointing towards a non-zero $w_c$. Results not shown here indicate that the time-evolving state's overlap with the PSS is itself an order parameter that shows transitions similar to $J^2$ and is consistent with the picture that before the transition, the state is largely in the PSS.}

{However, there is a qualitative difference if we start with a $J_y$ eigenstate such that every qubit is in the coherent state $\theta=\pi/2, \phi=\pi/2$. Here, the rotations about the $y$ axis have no impact on this state and the rotations about the $x-$axis keeps the states on the $y-z$ plane. We have shown that in the presence of disorder and $p=0$ these are exceptional and that the $J^2$
expectation value scales as $N$, even in the small $w$ case. Hence we are obtaining values of $w_c$ that seem to be flowing to a small value that may or may not be $0$, see Fig. \ref{diffic} (a). On the other hand the initial condition with all qubits starting from $\theta=0$, or the collective $J_z$ eigenstate has a larger value of $w_c$, although it is also on the exceptional great circle $\phi=\pi/2$. However, this initial state is strongly influenced by the rotation about the $y-$ axis and is immediately removed from this circle, and hence it behaves differently from the $J_y$ eigenstate.}

Returning to the generic initial states, for $w$ large and $O(1)$, the time-evolving states quickly leave the PSS and are generic superpositions of the basis states in any basis, 
the above approximation breaks down and numerical results indicate the formation of random states.
We have also confirmed the  generation of random states by studying the spectral level spacing statistics  which shows Wigner Dyson distribution at large disorder \cite{manju25}.
Indeed in this case the dynamics is similar to that of random circuits (for a recent review see \cite{RC_review2023}) that are known to generate psuedorandom quantum states that mimic true random states 
up to some order.
Studying systems as in this paper and related transitions from the point of view of such circuits for which there is a considerable current interest is promising.

It may also be useful to compare with the large body of work around many-body-localization, wherein strong disorder drives eigenstates into localized phases \cite{Vonneumannscaling,  MBL_pal, oganesyan_pal, MBL_Luitz}. In contrast to MBL, here we observe that the disorder breaks the permutation symmetry and takes the system to a chaotic dynamics in a larger Hilbert space dimension that leads to thermalisation of the system as opposed to localization in MBL. Although the canonical model of MBL, the Heisenberg 1D chain with a disordered onsite magnetic field \cite{MBL_pal} is a time-independent system, MBL has been noted in Floquet systems as well \cite{MBL_Floquet2016}. In fact the finite size scaling effects that plague such studies and make the phenomenon itself suspect seems to be milder for Floquet systems \cite{MBL_Floquet2023}. The system studied here differs crucially with all-to-all interactions, scaling of the disorder with the number of particles, and the existence of a classical limit in the permutation symmetric, disorder free case.
If there are further phase transitions in the present work, perhaps to a localized phase in the full Hilbert space remains to be seen. Despite the limitations of finite size systems, both MBL and the transitions observed in this work have clear relevance to the many-body systems being studied experimentally in this so-called NISQ (Noisy Intermediate-Scale Quantum) era.

\section{Conclusion}

{In this work, we studied the effect of introducing disorder in a system which can be described by collective angular momentum operators, where the dynamics is restricted within the permutation symmetric subspace of the total Hilbert space due to conservation of total angular momentum. The introduction of disorder breaks the permutation symmetry, and the dynamics explores the full Hilbert space where the expectation values of various observables, in the limit of large disorder, is given by the random matrix theory corresponding to the full Hilbert space. Using finite size scaling, we find a critical strength of disorder at which there is a second order phase transition from area law entangled states with total angular momentum scaling as $N$, to  volume law entangled states with total angular momentum scaling as $\sqrt{N}$.
}} We also observe a decrease in the critical disorder strength  as the chaos parameter $k$ is increased, pointing towards a zero critical strength for large $k$.
The exponents on the other hand seem to show a weaker dependence. 
But the scaling analysis performed on the finite size systems  shows relatively good collapse for $N=12,14,16,18$ along with the sharpening of the peak in the variance of $J^2$ at $w_c$ when the system size is increased, hinting towards possibility of a continuous phase transition in the thermodynamic limit.
Finally, while we do not underestimate the importance of finite sizes in these studies which may differ from results in the thermodynamic limit, experimental relevance to finite size systems is also noted.

From an applications perspective, our results show that for disorder less than the critical  strength, permutation symmetric systems are likely to be reliably used for their intended application, being at least not volume law entangled in the full Hilbert space. It is possible that the proposed phase transition be experimentally verified by extending few qubit experimental kicked top studies \cite{chaudhury_nature_husimi,Neill_2016,NMRstudiesin2qubit}, wherein disorder is anyway present. Among experimental platforms, trapped atomic ions and Rydberg atom arrays are widely used for realising interacting spin chains. Disorder in our system can be modeled in the form of site dependent laser fields, or by tuning the interactions that leads to variations in coupling strength \cite{trapped_ion_RevMod, quantum_ion_disordered, MBL_disorder_exp, Rydberg_exp}. Experimental realisation of kicked top \cite{Neill_2016, NMRstudiesin2qubit} is already done using the parameter regimes similar to those considered in our model, making an extension to disordered case experimentally feasible. In fact, a disordered all-to-all interacting system, simulating the SYK model, has been recently experimentally studied using a combination of atomic cloud in a cavity and controllable light shift in the large detuning limit \cite{sauerwein23}.

Among future directions, it will be interesting to explore if breaking permutation symmetry through other methods, say by introducing a power law interactions \cite{russomanno2021quantum}, will also exhibit  a similar phase transition as induced by disorder.

\section*{Data Availability}
All data generated or analysed during this study are included in this published article. Data are also available from the authors upon request. Please contact corresponding author Uma Divakaran for data.


\begin{thebibliography}{100}
\let\newblock\relax

\makeatletter
\newcommand{\JournalTitle}[1]{#1}
\newcommand{\bibinfo}[2]{#2}
\newcommand{\newblock}{}
\makeatother

\urlstyle{rm}
\expandafter\ifx\csname url\endcsname\relax
  \def\url#1{\texttt{#1}}\fi
\expandafter\ifx\csname urlprefix\endcsname\relax\def\urlprefix{URL }\fi
\expandafter\ifx\csname doiprefix\endcsname\relax\def\doiprefix{DOI: }\fi
\providecommand{\bibinfo}[2]{#2}
\providecommand{\eprint}[2][]{\url{#2}}

\bibitem{tabor1989chaos}
\bibinfo{author}{Tabor, M.}
\newblock \emph{\bibinfo{title}{Chaos and Integrability in Nonlinear Dynamics:
  An Introduction, .}} (\bibinfo{publisher}{WileyInterscience},
  \bibinfo{year}{1989}).

\bibitem{lichtenberg2013regular}
\bibinfo{author}{Lichtenberg, A.} \& \bibinfo{author}{Lieberman, M.}
\newblock \emph{\bibinfo{title}{Regular and Chaotic Dynamics}}.
\newblock Applied Mathematical Sciences (\bibinfo{publisher}{Springer New
  York}, \bibinfo{year}{2013}).

\bibitem{Ott_2002}
\bibinfo{author}{Ott, E.}
\newblock \emph{\bibinfo{title}{Chaos in Dynamical Systems}}
  (\bibinfo{publisher}{Cambridge University Press}, \bibinfo{year}{2002}),
  \bibinfo{edition}{2} edn.

\bibitem{gutzwiller1991chaos}
\bibinfo{author}{Gutzwiller, M.}
\newblock \emph{\bibinfo{title}{Chaos in Classical and Quantum Mechanics}}.
\newblock Interdisciplinary Applied Mathematics (\bibinfo{publisher}{Springer
  New York}, \bibinfo{year}{1991}).

\bibitem{Stockmann_1999}
\bibinfo{author}{Stöckmann, H.-J.}
\newblock \emph{\bibinfo{title}{Quantum Chaos: An Introduction}}
  (\bibinfo{publisher}{Cambridge University Press}, \bibinfo{year}{1999}).

\bibitem{mehta2004random}
\bibinfo{author}{Mehta, M.}
\newblock \emph{\bibinfo{title}{Random Matrices}}.
\newblock No. \bibinfo{number}{v. 142} in \bibinfo{series}{Pure and applied
  mathematics} (\bibinfo{publisher}{Elsevier/Academic Press},
  \bibinfo{year}{2004}).

\bibitem{Sandrobook}
\bibinfo{author}{Wimberger, S.}
\newblock \emph{\bibinfo{title}{Non Linear dynamics and Quantum Chaos - An
  Introduction}} (\bibinfo{publisher}{Springer International Publishing
  Switzerland}, \bibinfo{year}{2014}).

\bibitem{Rigol_polkonikov}
\bibinfo{author}{D'Alessio, L.}, \bibinfo{author}{Kafri, Y.},
  \bibinfo{author}{Polkovnikov, A.} \& \bibinfo{author}{Rigol, M.}
\newblock \bibinfo{journal}{\bibinfo{title}{From quantum chaos and eigenstate
  thermalization to statistical mechanics and thermodynamics}}.
\newblock {\emph{\JournalTitle{Advances in Physics}}}
  \textbf{\bibinfo{volume}{65}}, \bibinfo{pages}{239--362},
  \doiprefix\url{https://doi.org/10.1080/00018732.2016.1198134}
  (\bibinfo{year}{2016}).

\bibitem{kaufman2016quantum}
\bibinfo{author}{Kaufman, A.~M.} \emph{et~al.}
\newblock \bibinfo{journal}{\bibinfo{title}{Quantum thermalization through
  entanglement in an isolated many-body system}}.
\newblock {\emph{\JournalTitle{Science}}} \textbf{\bibinfo{volume}{353}},
  \bibinfo{pages}{794--800},
  \doiprefix\url{https://www.science.org/doi/10.1126/science.aaf6725}
  (\bibinfo{year}{2016}).

\bibitem{Bohigas}
\bibinfo{author}{Bohigas, O.}, \bibinfo{author}{Giannoni, M.~J.} \&
  \bibinfo{author}{Schmit, C.}
\newblock \bibinfo{journal}{\bibinfo{title}{Characterization of chaotic quantum
  spectra and universality of level fluctuation laws}}.
\newblock {\emph{\JournalTitle{Phys. Rev. Lett.}}}
  \textbf{\bibinfo{volume}{52}}, \bibinfo{pages}{1--4},
  \doiprefix\url{https://doi.org/10.1103/PhysRevLett.52.1}
  (\bibinfo{year}{1984}).

\bibitem{srednicki1994chaos}
\bibinfo{author}{Srednicki, M.}
\newblock \bibinfo{journal}{\bibinfo{title}{Chaos and quantum thermalization}}.
\newblock {\emph{\JournalTitle{Physical Review E}}}
  \textbf{\bibinfo{volume}{50}}, \bibinfo{pages}{888},
  \doiprefix\url{https://doi.org/10.1103/PhysRevE.50.888}
  (\bibinfo{year}{1994}).

\bibitem{griffiths69}
\bibinfo{author}{Griffiths, R.~B.}
\newblock \bibinfo{journal}{\bibinfo{title}{Nonanalytic behavior above the
  critical point in a random ising ferromagnet}}.
\newblock {\emph{\JournalTitle{Phys. Rev. Lett.}}}
  \textbf{\bibinfo{volume}{23}}, \bibinfo{pages}{17--19},
  \doiprefix\url{https://doi.org/10.1103/PhysRevLett.23.17}
  (\bibinfo{year}{1969}).

\bibitem{anderson58}
\bibinfo{author}{Anderson, P.~W.}
\newblock \bibinfo{journal}{\bibinfo{title}{Absence of diffusion in certain
  random lattices}}.
\newblock {\emph{\JournalTitle{Phys. Rev.}}} \textbf{\bibinfo{volume}{109}},
  \bibinfo{pages}{1492--1505},
  \doiprefix\url{https://doi.org/10.1103/PhysRev.109.1492}
  (\bibinfo{year}{1958}).

\bibitem{basko06}
\bibinfo{author}{Basko, D.}, \bibinfo{author}{Aleiner, I.} \&
  \bibinfo{author}{Altshuler, B.}
\newblock \bibinfo{journal}{\bibinfo{title}{Metal–insulator transition in a
  weakly interacting many-electron system with localized single-particle
  states}}.
\newblock {\emph{\JournalTitle{Annals of Physics}}}
  \textbf{\bibinfo{volume}{321}}, \bibinfo{pages}{1126--1205},
  \doiprefix\url{https://doi.org/10.1016/j.aop.2005.11.014}
  (\bibinfo{year}{2006}).

\bibitem{oganesyan_pal}
\bibinfo{author}{Oganesyan, V.} \& \bibinfo{author}{Huse, D.~A.}
\newblock \bibinfo{journal}{\bibinfo{title}{Localization of interacting
  fermions at high temperature}}.
\newblock {\emph{\JournalTitle{Phys. Rev. B}}} \textbf{\bibinfo{volume}{75}},
  \bibinfo{pages}{155111},
  \doiprefix\url{https://doi.org/10.1103/PhysRevB.75.155111}
  (\bibinfo{year}{2007}).

\bibitem{SYK_PRL_1993}
\bibinfo{author}{Sachdev, S.} \& \bibinfo{author}{Ye, J.}
\newblock \bibinfo{journal}{\bibinfo{title}{Gapless spin-fluid ground state in
  a random quantum heisenberg magnet}}.
\newblock {\emph{\JournalTitle{Phys. Rev. Lett.}}}
  \textbf{\bibinfo{volume}{70}}, \bibinfo{pages}{3339--3342},
  \doiprefix\url{https://doi.org/10.1103/PhysRevLett.70.3339}
  (\bibinfo{year}{1993}).

\bibitem{kitaev2015simple}
\bibinfo{author}{Kitaev, A.}
\newblock \bibinfo{title}{A simple model of quantum holography (part 2), 2015}.

\bibitem{maldacena2016remarks}
\bibinfo{author}{Maldacena, J.} \& \bibinfo{author}{Stanford, D.}
\newblock \bibinfo{journal}{\bibinfo{title}{Remarks on the sachdev-ye-kitaev
  model}}.
\newblock {\emph{\JournalTitle{Physical Review D}}}
  \textbf{\bibinfo{volume}{94}}, \bibinfo{pages}{106002},
  \doiprefix\url{https://doi.org/10.1103/PhysRevD.94.106002}
  (\bibinfo{year}{2016}).

\bibitem{grempel84}
\bibinfo{author}{Grempel, D.~R.}, \bibinfo{author}{Prange, R.~E.} \&
  \bibinfo{author}{Fishman, S.}
\newblock \bibinfo{journal}{\bibinfo{title}{Quantum dynamics of a nonintegrable
  system}}.
\newblock {\emph{\JournalTitle{Phys. Rev. A}}} \textbf{\bibinfo{volume}{29}},
  \bibinfo{pages}{1639--1647},
  \doiprefix\url{https://doi.org/10.1103/PhysRevA.29.1639}
  (\bibinfo{year}{1984}).

\bibitem{Devoret}
\bibinfo{author}{Koch, J.} \emph{et~al.}
\newblock \bibinfo{journal}{\bibinfo{title}{Charge-insensitive qubit design
  derived from the cooper pair box}}.
\newblock {\emph{\JournalTitle{Phys. Rev. A}}} \textbf{\bibinfo{volume}{76}},
  \bibinfo{pages}{042319},
  \doiprefix\url{https://doi.org/10.1103/PhysRevA.76.042319}
  (\bibinfo{year}{2007}).

\bibitem{PRL_Xmon}
\bibinfo{author}{Barends, R.} \emph{et~al.}
\newblock \bibinfo{journal}{\bibinfo{title}{Coherent josephson qubit suitable
  for scalable quantum integrated circuits}}.
\newblock {\emph{\JournalTitle{Phys. Rev. Lett.}}}
  \textbf{\bibinfo{volume}{111}}, \bibinfo{pages}{080502},
  \doiprefix\url{https://doi.org/10.1103/PhysRevLett.111.080502}
  (\bibinfo{year}{2013}).

\bibitem{transmon_latest}
\bibinfo{author}{Wang, C.} \emph{et~al.}
\newblock \bibinfo{journal}{\bibinfo{title}{Towards practical quantum
  computers: Transmon qubit with a lifetime approaching 0.5 milliseconds}}.
\newblock {\emph{\JournalTitle{npj Quantum Information}}}
  \textbf{\bibinfo{volume}{8}}, \bibinfo{pages}{3},
  \doiprefix\url{https://doi.org/10.1038/s41534-021-00510-2}
  (\bibinfo{year}{2022}).

\bibitem{Berke_2022}
\bibinfo{author}{Berke, C.}, \bibinfo{author}{Varvelis, E.},
  \bibinfo{author}{Trebst, S.}, \bibinfo{author}{Altland, A.} \&
  \bibinfo{author}{DiVincenzo, D.~P.}
\newblock \bibinfo{journal}{\bibinfo{title}{Transmon platform for quantum
  computing challenged by chaotic fluctuations}}.
\newblock {\emph{\JournalTitle{Nature Communications}}}
  \textbf{\bibinfo{volume}{13}}, \bibinfo{pages}{2495},
  \doiprefix\url{https://doi.org/10.1038/s41467-022-29940-y}
  (\bibinfo{year}{2022}).

\bibitem{zurek2006decoherence}
\bibinfo{author}{Zurek, W.}
\newblock \bibinfo{title}{Decoherence, chaos and the second law}.
\newblock In \emph{\bibinfo{booktitle}{Time And Matter}},
  \bibinfo{pages}{132--132},
  \doiprefix\url{https://doi.org/10.1103/PhysRevLett.72.2508}
  (\bibinfo{publisher}{World Scientific}, \bibinfo{year}{2006}).

\bibitem{trust_quantum_simulators}
\bibinfo{author}{Hauke, P.}, \bibinfo{author}{Cucchietti, F.~M.},
  \bibinfo{author}{Tagliacozzo, L.}, \bibinfo{author}{Deutsch, I.} \&
  \bibinfo{author}{Lewenstein, M.}
\newblock \bibinfo{journal}{\bibinfo{title}{Can one trust quantum simulators?}}
\newblock {\emph{\JournalTitle{Reports on Progress in Physics}}}
  \textbf{\bibinfo{volume}{75}}, \bibinfo{pages}{082401},
  \doiprefix\url{https://iopscience.iop.org/article/10.1088/0034-4885/75/8/082401/pdf}
  (\bibinfo{year}{2012}).

\bibitem{bandyopadhyay2002testing}
\bibinfo{author}{Bandyopadhyay, J.~N.} \& \bibinfo{author}{Lakshminarayan, A.}
\newblock \bibinfo{journal}{\bibinfo{title}{Testing statistical bounds on
  entanglement using quantum chaos}}.
\newblock {\emph{\JournalTitle{Physical Review Letters}}}
  \textbf{\bibinfo{volume}{89}}, \bibinfo{pages}{060402},
  \doiprefix\url{https://doi.org/10.1103/PhysRevLett.89.060402}
  (\bibinfo{year}{2002}).

\bibitem{gong2003intrinsic}
\bibinfo{author}{Gong, J.} \& \bibinfo{author}{Brumer, P.}
\newblock \bibinfo{journal}{\bibinfo{title}{Intrinsic decoherence dynamics in
  smooth hamiltonian systems: Quantum-classical correspondence}}.
\newblock {\emph{\JournalTitle{Physical Review A}}}
  \textbf{\bibinfo{volume}{68}}, \bibinfo{pages}{022101},
  \doiprefix\url{https://doi.org/10.1103/PhysRevA.68.022101}
  (\bibinfo{year}{2003}).

\bibitem{flambaum2000time}
\bibinfo{author}{Flambaum, V.~V.}
\newblock \bibinfo{journal}{\bibinfo{title}{Time dynamics in chaotic many-body
  systems: can chaos destroy a quantum computer?}}
\newblock {\emph{\JournalTitle{Australian Journal of Physics}}}
  \textbf{\bibinfo{volume}{53}}, \bibinfo{pages}{489--497},
  \doiprefix\url{https://doi.org/10.1071/PH99091} (\bibinfo{year}{2000}).

\bibitem{song2001quantum}
\bibinfo{author}{Song, P.~H.} \& \bibinfo{author}{Shepelyansky, D.~L.}
\newblock \bibinfo{journal}{\bibinfo{title}{Quantum computing of quantum chaos
  and imperfection effects}}.
\newblock {\emph{\JournalTitle{Physical Review Letters}}}
  \textbf{\bibinfo{volume}{86}}, \bibinfo{pages}{2162},
  \doiprefix\url{https://doi.org/10.1103/PhysRevLett.86.2162}
  (\bibinfo{year}{2001}).

\bibitem{georgeot2000emergence}
\bibinfo{author}{Georgeot, B.} \& \bibinfo{author}{Shepelyansky, D.~L.}
\newblock \bibinfo{journal}{\bibinfo{title}{Emergence of quantum chaos in the
  quantum computer core and how to manage it}}.
\newblock {\emph{\JournalTitle{Physical Review E}}}
  \textbf{\bibinfo{volume}{62}}, \bibinfo{pages}{6366},
  \doiprefix\url{https://doi.org/10.1103/PhysRevE.62.6366}
  (\bibinfo{year}{2000}).

\bibitem{braun2002quantum}
\bibinfo{author}{Braun, D.}
\newblock \bibinfo{journal}{\bibinfo{title}{Quantum chaos and quantum
  algorithms}}.
\newblock {\emph{\JournalTitle{Physical Review A}}}
  \textbf{\bibinfo{volume}{65}}, \bibinfo{pages}{042317},
  \doiprefix\url{https://doi.org/10.1103/PhysRevA.65.042317}
  (\bibinfo{year}{2002}).

\bibitem{madhok2018quantum}
\bibinfo{author}{Madhok, V.}, \bibinfo{author}{Dogra, S.} \&
  \bibinfo{author}{Lakshminarayan, A.}
\newblock \bibinfo{journal}{\bibinfo{title}{Quantum correlations as probes of
  chaos and ergodicity}}.
\newblock {\emph{\JournalTitle{Optics Communications}}}
  \textbf{\bibinfo{volume}{420}}, \bibinfo{pages}{189--193},
  \doiprefix\url{https://doi.org/10.1016/j.optcom.2018.03.069}
  (\bibinfo{year}{2018}).

\bibitem{chaos_and_computers}
\bibinfo{author}{Shepelyansky, D.}
\newblock \bibinfo{journal}{\bibinfo{title}{Quantum chaos and quantum
  computers}}.
\newblock {\emph{\JournalTitle{Physica Scripta}}}
  \textbf{\bibinfo{volume}{2001}}, \bibinfo{pages}{112},
  \doiprefix\url{https://iopscience.iop.org/article/10.1238/Physica.Topical.090a00112/pdf}
  (\bibinfo{year}{2001}).

\bibitem{piga2019quantum}
\bibinfo{author}{Piga, A.}, \bibinfo{author}{Lewenstein, M.} \&
  \bibinfo{author}{Quach, J.~Q.}
\newblock \bibinfo{journal}{\bibinfo{title}{Quantum chaos and entanglement in
  ergodic and nonergodic systems}}.
\newblock {\emph{\JournalTitle{Physical Review E}}}
  \textbf{\bibinfo{volume}{99}}, \bibinfo{pages}{032213},
  \doiprefix\url{https://doi.org/10.1103/PhysRevE.99.032213}
  (\bibinfo{year}{2019}).

\bibitem{LMG_PRR}
\bibinfo{author}{Chinni, K.}, \bibinfo{author}{Poggi, P.~M.} \&
  \bibinfo{author}{Deutsch, I.~H.}
\newblock \bibinfo{journal}{\bibinfo{title}{Effect of chaos on the simulation
  of quantum critical phenomena in analog quantum simulators}}.
\newblock {\emph{\JournalTitle{Physical Review Research}}}
  \textbf{\bibinfo{volume}{3}}, \bibinfo{pages}{033145},
  \doiprefix\url{https://doi.org/10.1103/PhysRevResearch.3.033145}
  (\bibinfo{year}{2021}).

\bibitem{papparaldi_bridging}
\bibinfo{author}{Lerose, A.} \& \bibinfo{author}{Pappalardi, S.}
\newblock \bibinfo{journal}{\bibinfo{title}{Bridging entanglement dynamics and
  chaos in semiclassical systems}}.
\newblock {\emph{\JournalTitle{Physical Review A}}}
  \textbf{\bibinfo{volume}{102}}, \bibinfo{pages}{032404},
  \doiprefix\url{https://doi.org/10.1103/PhysRevA.102.032404}
  (\bibinfo{year}{2020}).

\bibitem{taming_disorder}
\bibinfo{author}{Braiman, Y.}, \bibinfo{author}{Lindner, J.~F.} \&
  \bibinfo{author}{Ditto, W.~L.}
\newblock \bibinfo{journal}{\bibinfo{title}{Taming spatiotemporal chaos with
  disorder}}.
\newblock {\emph{\JournalTitle{Nature}}} \textbf{\bibinfo{volume}{378}},
  \bibinfo{pages}{465--467}, \doiprefix\url{https://doi.org/10.1038/378465a0}
  (\bibinfo{year}{1995}).

\bibitem{Vonneumannscaling}
\bibinfo{author}{Lee, M.}, \bibinfo{author}{Look, T.~R.}, \bibinfo{author}{Lim,
  S.~P.} \& \bibinfo{author}{Sheng, D.~N.}
\newblock \bibinfo{journal}{\bibinfo{title}{Many-body localization in spin
  chain systems with quasiperiodic fields}}.
\newblock {\emph{\JournalTitle{Phys. Rev. B}}} \textbf{\bibinfo{volume}{96}},
  \bibinfo{pages}{075146},
  \doiprefix\url{https://doi.org/10.1103/PhysRevB.96.075146}
  (\bibinfo{year}{2017}).

\bibitem{MBL_pal}
\bibinfo{author}{Pal, A.} \& \bibinfo{author}{Huse, D.~A.}
\newblock \bibinfo{journal}{\bibinfo{title}{Many-body localization phase
  transition}}.
\newblock {\emph{\JournalTitle{Phys. Rev. B}}} \textbf{\bibinfo{volume}{82}},
  \bibinfo{pages}{174411},
  \doiprefix\url{https://doi.org/10.1103/PhysRevB.82.174411}
  (\bibinfo{year}{2010}).

\bibitem{MBL_Luitz}
\bibinfo{author}{Luitz, D.~J.}, \bibinfo{author}{Laflorencie, N.} \&
  \bibinfo{author}{Alet, F.}
\newblock \bibinfo{journal}{\bibinfo{title}{Many-body localization edge in the
  random-field heisenberg chain}}.
\newblock {\emph{\JournalTitle{Phys. Rev. B}}} \textbf{\bibinfo{volume}{91}},
  \bibinfo{pages}{081103},
  \doiprefix\url{https://doi.org/10.1103/PhysRevB.91.081103}
  (\bibinfo{year}{2015}).

\bibitem{tripartitescrambling}
\bibinfo{author}{Seshadri, A.}, \bibinfo{author}{Madhok, V.} \&
  \bibinfo{author}{Lakshminarayan, A.}
\newblock \bibinfo{journal}{\bibinfo{title}{Tripartite mutual information,
  entanglement, and scrambling in permutation symmetric systems with an
  application to quantum chaos}}.
\newblock {\emph{\JournalTitle{Physical Review E}}}
  \textbf{\bibinfo{volume}{98}}, \bibinfo{pages}{052205},
  \doiprefix\url{https://doi.org/10.1103/PhysRevE.98.052205}
  (\bibinfo{year}{2018}).

\bibitem{scully_single_photon}
\bibinfo{journal}{\bibinfo{author}{Scully, M.~O.}}
\newblock {\emph{\JournalTitle{Physical Review Letters}}}
  \textbf{\bibinfo{volume}{115}}, \bibinfo{pages}{243602}
  (\bibinfo{year}{2015}).

\bibitem{J.physicssuperradiant}
\bibinfo{author}{Gegg, M.}, \bibinfo{author}{Carmele, A.},
  \bibinfo{author}{Knorr, A.} \& \bibinfo{author}{Richter, M.}
\newblock \bibinfo{journal}{\bibinfo{title}{Superradiant to subradiant phase
  transition in the open system dicke model: Dark state cascades}}.
\newblock {\emph{\JournalTitle{New Journal of Physics}}}
  \textbf{\bibinfo{volume}{20}}, \bibinfo{pages}{013006},
  \doiprefix\url{https://iopscience.iop.org/article/10.1088/1367-2630/aa9cdd}
  (\bibinfo{year}{2018}).

\bibitem{rubies2022superradiance}
\bibinfo{author}{Rubies-Bigorda, O.} \& \bibinfo{author}{Yelin, S.~F.}
\newblock \bibinfo{journal}{\bibinfo{title}{Superradiance and subradiance in
  inverted atomic arrays}}.
\newblock {\emph{\JournalTitle{Physical Review A}}}
  \textbf{\bibinfo{volume}{106}}, \bibinfo{pages}{053717},
  \doiprefix\url{https://doi.org/10.1103/PhysRevA.106.053717}
  (\bibinfo{year}{2022}).

\bibitem{DHS_23}
\bibinfo{author}{Qi, Z.}, \bibinfo{author}{Scaffidi, T.} \&
  \bibinfo{author}{Cao, X.}
\newblock \bibinfo{journal}{\bibinfo{title}{Surprises in the deep hilbert space
  of all-to-all systems: From superexponential scrambling to slow entanglement
  growth}}.
\newblock {\emph{\JournalTitle{Phys. Rev. B}}} \textbf{\bibinfo{volume}{108}},
  \bibinfo{pages}{054301},
  \doiprefix\url{https://doi.org/10.1103/PhysRevB.108.054301}
  (\bibinfo{year}{2023}).

\bibitem{iemini2024dynamics}
\bibinfo{author}{Iemini, F.}, \bibinfo{author}{Chang, D.} \&
  \bibinfo{author}{Marino, J.}
\newblock \bibinfo{journal}{\bibinfo{title}{Dynamics of inhomogeneous spin
  ensembles with all-to-all interactions: Breaking permutational invariance}}.
\newblock {\emph{\JournalTitle{Phys. Rev. A}}} \textbf{\bibinfo{volume}{109}},
  \bibinfo{pages}{032204},
  \doiprefix\url{https://doi.org/10.1103/PhysRevA.109.032204}
  (\bibinfo{year}{2024}).

\bibitem{haake1987classical}
\bibinfo{author}{Haake, F.}, \bibinfo{author}{Ku{\'s}, M.} \&
  \bibinfo{author}{Scharf, R.}
\newblock \bibinfo{journal}{\bibinfo{title}{Classical and quantum chaos for a
  kicked top}}.
\newblock {\emph{\JournalTitle{Zeitschrift f{\"u}r Physik B Condensed Matter}}}
  \textbf{\bibinfo{volume}{65}}, \bibinfo{pages}{381--395},
  \doiprefix\url{https://doi.org/10.1007/BF01303727} (\bibinfo{year}{1987}).

\bibitem{PhysRevA.45.3646}
\bibinfo{author}{D'Ariano, G.~M.}, \bibinfo{author}{Evangelista, L.~R.} \&
  \bibinfo{author}{Saraceno, M.}
\newblock \bibinfo{journal}{\bibinfo{title}{Classical and quantum structures in
  the kicked-top model}}.
\newblock {\emph{\JournalTitle{Phys. Rev. A}}} \textbf{\bibinfo{volume}{45}},
  \bibinfo{pages}{3646--3658},
  \doiprefix\url{https://doi.org/10.1103/PhysRevA.45.3646}
  (\bibinfo{year}{1992}).

\bibitem{Haakebook}
\bibinfo{author}{Haake, F.}, \bibinfo{author}{Gnutzmann, S.} \&
  \bibinfo{author}{Ku{\'s}, M.}
\newblock \emph{\bibinfo{title}{Quantum Signatures of Chaos}}.
\newblock Springer complexity (\bibinfo{publisher}{Springer},
  \bibinfo{year}{2018}).

\bibitem{concurrence}
\bibinfo{author}{Wang, X.}, \bibinfo{author}{Ghose, S.},
  \bibinfo{author}{Sanders, B.~C.} \& \bibinfo{author}{Hu, B.}
\newblock \bibinfo{journal}{\bibinfo{title}{Entanglement as a signature of
  quantum chaos}}.
\newblock {\emph{\JournalTitle{Physical Review E}}}
  \textbf{\bibinfo{volume}{70}}, \bibinfo{pages}{016217},
  \doiprefix\url{https://doi.org/10.1103/PhysRevE.70.016217}
  (\bibinfo{year}{2004}).

\bibitem{madhok_quantum_discord}
\bibinfo{author}{Madhok, V.}, \bibinfo{author}{Gupta, V.},
  \bibinfo{author}{Trottier, D.-A.} \& \bibinfo{author}{Ghose, S.}
\newblock \bibinfo{journal}{\bibinfo{title}{Signatures of chaos in the dynamics
  of quantum discord}}.
\newblock {\emph{\JournalTitle{Physical Review E}}}
  \textbf{\bibinfo{volume}{91}}, \bibinfo{pages}{032906},
  \doiprefix\url{https://doi.org/10.1103/PhysRevE.91.032906}
  (\bibinfo{year}{2015}).

\bibitem{chaudhury_nature_husimi}
\bibinfo{author}{Chaudhury, S.}, \bibinfo{author}{Smith, A.},
  \bibinfo{author}{Anderson, B.}, \bibinfo{author}{Ghose, S.} \&
  \bibinfo{author}{Jessen, P.~S.}
\newblock \bibinfo{journal}{\bibinfo{title}{Quantum signatures of chaos in a
  kicked top}}.
\newblock {\emph{\JournalTitle{Nature}}} \textbf{\bibinfo{volume}{461}},
  \bibinfo{pages}{768--771},
  \doiprefix\url{https://doi.org/10.1038/nature08396} (\bibinfo{year}{2009}).

\bibitem{Neill_2016}
\bibinfo{author}{Neill, C.} \emph{et~al.}
\newblock \bibinfo{journal}{\bibinfo{title}{Ergodic dynamics and thermalization
  in an isolated quantum system}}.
\newblock {\emph{\JournalTitle{Nature Physics}}} \textbf{\bibinfo{volume}{12}},
  \bibinfo{pages}{1037}, \doiprefix\url{https://doi.org/10.1038/nphys3830}
  (\bibinfo{year}{2016}).

\bibitem{NMRstudiesin2qubit}
\bibinfo{author}{Krithika, V.}, \bibinfo{author}{Anjusha, V.},
  \bibinfo{author}{Bhosale, U.~T.} \& \bibinfo{author}{Mahesh, T.}
\newblock \bibinfo{journal}{\bibinfo{title}{Nmr studies of quantum chaos in a
  two-qubit kicked top}}.
\newblock {\emph{\JournalTitle{Physical Review E}}}
  \textbf{\bibinfo{volume}{99}}, \bibinfo{pages}{032219},
  \doiprefix\url{https://doi.org/10.1103/PhysRevE.99.032219}
  (\bibinfo{year}{2019}).

\bibitem{zyczkowski1990indicators}
\bibinfo{author}{Zyczkowski, K.}
\newblock \bibinfo{journal}{\bibinfo{title}{Indicators of quantum chaos based
  on eigenvector statistics}}.
\newblock {\emph{\JournalTitle{Journal of Physics A: Mathematical and
  General}}} \textbf{\bibinfo{volume}{23}}, \bibinfo{pages}{4427},
  \doiprefix\url{10.1088/0305-4470/23/20/005} (\bibinfo{year}{1990}).

\bibitem{effect_measurement_qkt}
\bibinfo{author}{Sanders, B.} \& \bibinfo{author}{Milburn, G.}
\newblock \bibinfo{journal}{\bibinfo{title}{The effect of measurement on the
  quantum features of a chaotic system}}.
\newblock {\emph{\JournalTitle{Zeitschrift f{\"u}r Physik B Condensed Matter}}}
  \textbf{\bibinfo{volume}{77}}, \bibinfo{pages}{497--510},
  \doiprefix\url{https://doi.org/10.1007/BF01453801} (\bibinfo{year}{1989}).

\bibitem{periodicity}
\bibinfo{author}{Bhosale, U.~T.} \& \bibinfo{author}{Santhanam, M.}
\newblock \bibinfo{journal}{\bibinfo{title}{Periodicity of quantum correlations
  in the quantum kicked top}}.
\newblock {\emph{\JournalTitle{Physical Review E}}}
  \textbf{\bibinfo{volume}{98}}, \bibinfo{pages}{052228},
  \doiprefix\url{https://doi.org/10.1103/PhysRevE.98.052228}
  (\bibinfo{year}{2018}).

\bibitem{fox1994chaos}
\bibinfo{author}{Fox, R.~F.} \& \bibinfo{author}{Elston, T.~C.}
\newblock \bibinfo{journal}{\bibinfo{title}{Chaos and a quantum-classical
  correspondence in the kicked top}}.
\newblock {\emph{\JournalTitle{Physical Review E}}}
  \textbf{\bibinfo{volume}{50}}, \bibinfo{pages}{2553},
  \doiprefix\url{https://doi.org/10.1103/PhysRevE.50.2553}
  (\bibinfo{year}{1994}).

\bibitem{spinsqueezing}
\bibinfo{author}{Wang, X.}, \bibinfo{author}{Ma, J.}, \bibinfo{author}{Song,
  L.}, \bibinfo{author}{Zhang, X.} \& \bibinfo{author}{Wang, X.}
\newblock \bibinfo{journal}{\bibinfo{title}{Spin squeezing, negative
  correlations, and concurrence in the quantum kicked top model}}.
\newblock {\emph{\JournalTitle{Physical Review E}}}
  \textbf{\bibinfo{volume}{82}}, \bibinfo{pages}{056205},
  \doiprefix\url{https://doi.org/10.1103/PhysRevE.82.056205}
  (\bibinfo{year}{2010}).

\bibitem{zou2022pseudoclassical}
\bibinfo{author}{Zou, Z.} \& \bibinfo{author}{Wang, J.}
\newblock \bibinfo{journal}{\bibinfo{title}{Pseudoclassical dynamics of the
  kicked top}}.
\newblock {\emph{\JournalTitle{Entropy}}} \textbf{\bibinfo{volume}{24}},
  \bibinfo{pages}{1092}, \doiprefix\url{https://doi.org/10.3390/e24081092}
  (\bibinfo{year}{2022}).

\bibitem{kickedpspin}
\bibinfo{author}{Mu{\~n}oz-Arias, M.~H.}, \bibinfo{author}{Poggi, P.~M.} \&
  \bibinfo{author}{Deutsch, I.~H.}
\newblock \bibinfo{journal}{\bibinfo{title}{Nonlinear dynamics and quantum
  chaos of a family of kicked p-spin models}}.
\newblock {\emph{\JournalTitle{Physical Review E}}}
  \textbf{\bibinfo{volume}{103}}, \bibinfo{pages}{052212},
  \doiprefix\url{https://doi.org/10.1103/PhysRevE.103.052212}
  (\bibinfo{year}{2021}).

\bibitem{quantummetrology}
\bibinfo{author}{Fiderer, L.~J.} \& \bibinfo{author}{Braun, D.}
\newblock \bibinfo{journal}{\bibinfo{title}{Quantum metrology with
  quantum-chaotic sensors}}.
\newblock {\emph{\JournalTitle{Nature Communications}}}
  \textbf{\bibinfo{volume}{9}}, \bibinfo{pages}{1351},
  \doiprefix\url{https://doi.org/10.1038/s41467-018-03623-z}
  (\bibinfo{year}{2018}).

\bibitem{bifurcation_QKT}
\bibinfo{author}{Bhosale, U.~T.} \& \bibinfo{author}{Santhanam, M.}
\newblock \bibinfo{journal}{\bibinfo{title}{Signatures of bifurcation on
  quantum correlations: Case of the quantum kicked top}}.
\newblock {\emph{\JournalTitle{Physical Review E}}}
  \textbf{\bibinfo{volume}{95}}, \bibinfo{pages}{012216},
  \doiprefix\url{https://doi.org/10.1103/PhysRevE.95.012216}
  (\bibinfo{year}{2017}).

\bibitem{sieberer2019digital}
\bibinfo{author}{Sieberer, L.~M.} \emph{et~al.}
\newblock \bibinfo{journal}{\bibinfo{title}{Digital quantum simulation, trotter
  errors, and quantum chaos of the kicked top}}.
\newblock {\emph{\JournalTitle{npj Quantum Information}}}
  \textbf{\bibinfo{volume}{5}}, \bibinfo{pages}{78},
  \doiprefix\url{https://doi.org/10.1038/s41534-019-0192-5}
  (\bibinfo{year}{2019}).

\bibitem{periodic_orbits}
\bibinfo{author}{Kumari, M.} \& \bibinfo{author}{Ghose, S.}
\newblock \bibinfo{journal}{\bibinfo{title}{Quantum-classical correspondence in
  the vicinity of periodic orbits}}.
\newblock {\emph{\JournalTitle{Phys. Rev. E}}} \textbf{\bibinfo{volume}{97}},
  \bibinfo{pages}{052209},
  \doiprefix\url{https://doi.org/10.1103/PhysRevE.97.052209}
  (\bibinfo{year}{2018}).

\bibitem{lombardi2011entanglement}
\bibinfo{author}{Lombardi, M.} \& \bibinfo{author}{Matzkin, A.}
\newblock \bibinfo{journal}{\bibinfo{title}{Entanglement and chaos in the
  kicked top}}.
\newblock {\emph{\JournalTitle{Physical Review E}}}
  \textbf{\bibinfo{volume}{83}}, \bibinfo{pages}{016207},
  \doiprefix\url{https://doi.org/10.1103/PhysRevE.83.016207}
  (\bibinfo{year}{2011}).

\bibitem{Amit_Anand_2023}
\bibinfo{author}{Anand, A.}, \bibinfo{author}{Davis, J.} \&
  \bibinfo{author}{Ghose, S.}
\newblock \bibinfo{journal}{\bibinfo{title}{Quantum recurrences in the kicked
  top}}.
\newblock {\emph{\JournalTitle{Phys. Rev. Res.}}} \textbf{\bibinfo{volume}{6}},
  \bibinfo{pages}{023120},
  \doiprefix\url{https://doi.org/10.1103/PhysRevResearch.6.023120}
  (\bibinfo{year}{2024}).

\bibitem{decoherence}
\bibinfo{author}{Ghose, S.}, \bibinfo{author}{Stock, R.},
  \bibinfo{author}{Jessen, P.}, \bibinfo{author}{Lal, R.} \&
  \bibinfo{author}{Silberfarb, A.}
\newblock \bibinfo{journal}{\bibinfo{title}{Chaos, entanglement, and
  decoherence in the quantum kicked top}}.
\newblock {\emph{\JournalTitle{Physical Review A}}}
  \textbf{\bibinfo{volume}{78}}, \bibinfo{pages}{042318},
  \doiprefix\url{https://doi.org/10.1103/PhysRevA.78.042318}
  (\bibinfo{year}{2008}).

\bibitem{defenurmp}
\bibinfo{author}{Defenu, N.} \emph{et~al.}
\newblock \bibinfo{journal}{\bibinfo{title}{Long-range interacting quantum
  systems}}.
\newblock {\emph{\JournalTitle{Rev. Mod. Phys.}}}
  \textbf{\bibinfo{volume}{95}}, \bibinfo{pages}{035002},
  \doiprefix\url{https://doi.org/10.1103/RevModPhys.95.035002}
  (\bibinfo{year}{2023}).

\bibitem{defunu24}
\bibinfo{author}{Defenu, N.}, \bibinfo{author}{Lerose, A.} \&
  \bibinfo{author}{Pappalardi, S.}
\newblock \bibinfo{journal}{\bibinfo{title}{Out-of-equilibrium dynamics of
  quantum many-body systems with long-range interactions}}.
\newblock {\emph{\JournalTitle{Physics Reports}}}
  \textbf{\bibinfo{volume}{1074}}, \bibinfo{pages}{1--92},
  \doiprefix\url{https://doi.org/10.1016/j.physrep.2024.04.005}
  (\bibinfo{year}{2024}).

\bibitem{pappalardi2018scrambling}
\bibinfo{author}{Pappalardi, S.} \emph{et~al.}
\newblock \bibinfo{journal}{\bibinfo{title}{Scrambling and entanglement
  spreading in long-range spin chains}}.
\newblock {\emph{\JournalTitle{Physical Review B}}}
  \textbf{\bibinfo{volume}{98}}, \bibinfo{pages}{134303},
  \doiprefix\url{https://doi.org/10.1103/PhysRevB.98.134303}
  (\bibinfo{year}{2018}).

\bibitem{constantoudis1997lyapunov}
\bibinfo{author}{Constantoudis, V.} \& \bibinfo{author}{Theodorakopoulos, N.}
\newblock \bibinfo{journal}{\bibinfo{title}{Lyapunov exponent, stretching
  numbers, and islands of stability of the kicked top}}.
\newblock {\emph{\JournalTitle{Physical Review E}}}
  \textbf{\bibinfo{volume}{56}}, \bibinfo{pages}{5189},
  \doiprefix\url{https://doi.org/10.1103/PhysRevE.56.5189}
  (\bibinfo{year}{1997}).

\bibitem{wang2021multifractality}
\bibinfo{author}{Wang, Q.} \& \bibinfo{author}{Robnik, M.}
\newblock \bibinfo{journal}{\bibinfo{title}{Multifractality in quasienergy
  space of coherent states as a signature of quantum chaos}}.
\newblock {\emph{\JournalTitle{Entropy}}} \textbf{\bibinfo{volume}{23}},
  \bibinfo{pages}{1347}, \doiprefix\url{https://doi.org/10.3390/e23101347}
  (\bibinfo{year}{2021}).

\bibitem{glauber_spin_coherent_state}
\bibinfo{author}{Glauber, R.~J.} \& \bibinfo{author}{Haake, F.}
\newblock \bibinfo{journal}{\bibinfo{title}{Superradiant pulses and directed
  angular momentum states}}.
\newblock {\emph{\JournalTitle{Physical Review A}}}
  \textbf{\bibinfo{volume}{13}}, \bibinfo{pages}{357},
  \doiprefix\url{https://doi.org/10.1103/PhysRevA.13.357}
  (\bibinfo{year}{1976}).

\bibitem{spincoherentstates}
\bibinfo{author}{Lee~Loh, Y.} \& \bibinfo{author}{Kim, M.}
\newblock \bibinfo{journal}{\bibinfo{title}{Visualizing spin states using the
  spin coherent state representation}}.
\newblock {\emph{\JournalTitle{American Journal of Physics}}}
  \textbf{\bibinfo{volume}{83}}, \bibinfo{pages}{30--35},
  \doiprefix\url{https://doi.org/10.1119/1.4898595} (\bibinfo{year}{2015}).

\bibitem{R.R_puri}
\bibinfo{author}{Puri, R.}
\newblock \emph{\bibinfo{title}{Mathematical Methods of Quantum Optics}}
  (\bibinfo{publisher}{Springer Berlin}, \bibinfo{year}{2001}).

\bibitem{pattnayak}
\bibinfo{author}{Ruebeck, J.~B.}, \bibinfo{author}{Lin, J.} \&
  \bibinfo{author}{Pattanayak, A.~K.}
\newblock \bibinfo{journal}{\bibinfo{title}{Entanglement and its relationship
  to classical dynamics}}.
\newblock {\emph{\JournalTitle{Physical Review E}}}
  \textbf{\bibinfo{volume}{95}}, \bibinfo{pages}{062222},
  \doiprefix\url{https://doi.org/10.1103/PhysRevE.95.062222}
  (\bibinfo{year}{2017}).

\bibitem{fewbodykickedtop}
\bibinfo{author}{Dogra, S.}, \bibinfo{author}{Madhok, V.} \&
  \bibinfo{author}{Lakshminarayan, A.}
\newblock \bibinfo{journal}{\bibinfo{title}{Quantum signatures of chaos,
  thermalization, and tunneling in the exactly solvable few-body kicked top}}.
\newblock {\emph{\JournalTitle{Physical Review E}}}
  \textbf{\bibinfo{volume}{99}}, \bibinfo{pages}{062217},
  \doiprefix\url{https://doi.org/10.1103/PhysRevE.99.062217}
  (\bibinfo{year}{2019}).

\bibitem{otocandloschmidt}
\bibinfo{author}{Sreeram, P.}, \bibinfo{author}{Madhok, V.} \&
  \bibinfo{author}{Lakshminarayan, A.}
\newblock \bibinfo{journal}{\bibinfo{title}{Out-of-time-ordered correlators and
  the loschmidt echo in the quantum kicked top: how low can we go?}}
\newblock {\emph{\JournalTitle{Journal of Physics D: Applied Physics}}}
  \textbf{\bibinfo{volume}{54}}, \bibinfo{pages}{274004},
  \doiprefix\url{https://iopscience.iop.org/article/10.1088/1361-6463/abf8f3/pdf}
  (\bibinfo{year}{2021}).

\bibitem{castro2013entanglement}
\bibinfo{author}{Castro-Alvaredo, O.~A.} \& \bibinfo{author}{Doyon, B.}
\newblock \bibinfo{journal}{\bibinfo{title}{Entanglement in permutation
  symmetric states, fractal dimensions, and geometric quantum mechanics}}.
\newblock {\emph{\JournalTitle{Journal of Statistical Mechanics: Theory and
  Experiment}}} \textbf{\bibinfo{volume}{2013}}, \bibinfo{pages}{P02016},
  \doiprefix\url{https://iopscience.iop.org/article/10.1088/1742-5468/2013/02/P02016/pdf}
  (\bibinfo{year}{2013}).

\bibitem{page1993average}
\bibinfo{author}{Page, D.~N.}
\newblock \bibinfo{journal}{\bibinfo{title}{Average entropy of a subsystem}}.
\newblock {\emph{\JournalTitle{Physical Review Letters}}}
  \textbf{\bibinfo{volume}{71}}, \bibinfo{pages}{1291},
  \doiprefix\url{https://doi.org/10.1103/PhysRevLett.71.1291}
  (\bibinfo{year}{1993}).

\bibitem{sherington_model}
\bibinfo{author}{Young, A.}
\newblock \bibinfo{journal}{\bibinfo{title}{Stability of the quantum
  sherrington-kirkpatrick spin glass model}}.
\newblock {\emph{\JournalTitle{Physical Review E}}}
  \textbf{\bibinfo{volume}{96}}, \bibinfo{pages}{032112},
  \doiprefix\url{https://doi.org/10.1103/PhysRevE.96.032112}
  (\bibinfo{year}{2017}).

\bibitem{kirkpatrick_infinite_range}
\bibinfo{author}{Thirumalai, D.}, \bibinfo{author}{Li, Q.} \&
  \bibinfo{author}{Kirkpatrick, T.}
\newblock \bibinfo{journal}{\bibinfo{title}{Infinite-range ising spin glass in
  a transverse field}}.
\newblock {\emph{\JournalTitle{Journal of Physics A: Mathematical and
  General}}} \textbf{\bibinfo{volume}{22}}, \bibinfo{pages}{3339},
  \doiprefix\url{https://iopscience.iop.org/article/10.1088/0305-4470/22/16/023/pdf}
  (\bibinfo{year}{1989}).

\bibitem{silvia_sk}
\bibinfo{author}{Pappalardi, S.}, \bibinfo{author}{Polkovnikov, A.} \&
  \bibinfo{author}{Silva, A.}
\newblock \bibinfo{journal}{\bibinfo{title}{{Quantum echo dynamics in the
  Sherrington-Kirkpatrick model}}}.
\newblock {\emph{\JournalTitle{SciPost Phys.}}} \textbf{\bibinfo{volume}{9}},
  \bibinfo{pages}{021},
  \doiprefix\url{https://scipost.org/10.21468/SciPostPhys.9.2.021}
  (\bibinfo{year}{2020}).

\bibitem{botetpfeuty82}
\bibinfo{author}{Botet, R.}, \bibinfo{author}{Jullien, R.} \&
  \bibinfo{author}{Pfeuty, P.}
\newblock \bibinfo{journal}{\bibinfo{title}{Size scaling for infinitely
  coordinated systems}}.
\newblock {\emph{\JournalTitle{Phys. Rev. Lett.}}}
  \textbf{\bibinfo{volume}{49}}, \bibinfo{pages}{478--481},
  \doiprefix\url{https://doi.org/10.1103/PhysRevB.28.3955}
  (\bibinfo{year}{1982}).

\bibitem{botet83}
\bibinfo{author}{Botet, R.} \& \bibinfo{author}{Jullien, R.}
\newblock \bibinfo{journal}{\bibinfo{title}{Large-size critical behavior of
  infinitely coordinated systems}}.
\newblock {\emph{\JournalTitle{Phys. Rev. B}}} \textbf{\bibinfo{volume}{28}},
  \bibinfo{pages}{3955--3967}, \doiprefix\url{10.1103/PhysRevB.28.3955}
  (\bibinfo{year}{1983}).

\bibitem{curveJ2similar}
\bibinfo{author}{Skinner, B.}, \bibinfo{author}{Ruhman, J.} \&
  \bibinfo{author}{Nahum, A.}
\newblock \bibinfo{journal}{\bibinfo{title}{Measurement-induced phase
  transitions in the dynamics of entanglement}}.
\newblock {\emph{\JournalTitle{Phys. Rev. X}}} \textbf{\bibinfo{volume}{9}},
  \bibinfo{pages}{031009},
  \doiprefix\url{https://doi.org/10.1103/PhysRevX.9.031009}
  (\bibinfo{year}{2019}).

\bibitem{dynamicalcriticalscaling}
\bibinfo{author}{Li, Y.}, \bibinfo{author}{Chen, X.} \&
  \bibinfo{author}{Fisher, M. P.~A.}
\newblock \bibinfo{journal}{\bibinfo{title}{Quantum zeno effect and the
  many-body entanglement transition}}.
\newblock {\emph{\JournalTitle{Phys. Rev. B}}} \textbf{\bibinfo{volume}{98}},
  \bibinfo{pages}{205136},
  \doiprefix\url{https://doi.org/10.1103/PhysRevB.98.205136}
  (\bibinfo{year}{2018}).

\bibitem{cardy1988finite}
\bibinfo{author}{Cardy, J.}
\newblock \emph{\bibinfo{title}{Finite-size Scaling}}.
\newblock Current physics (\bibinfo{publisher}{North-Holland},
  \bibinfo{year}{1988}).

\bibitem{khemani2017critical}
\bibinfo{author}{Khemani, V.}, \bibinfo{author}{Lim, S.-P.},
  \bibinfo{author}{Sheng, D.} \& \bibinfo{author}{Huse, D.~A.}
\newblock \bibinfo{journal}{\bibinfo{title}{Critical properties of the
  many-body localization transition}}.
\newblock {\emph{\JournalTitle{Physical Review X}}}
  \textbf{\bibinfo{volume}{7}}, \bibinfo{pages}{021013},
  \doiprefix\url{https://doi.org/10.1103/PhysRevX.7.021013}
  (\bibinfo{year}{2017}).

\bibitem{abanin2021distinguishing}
\bibinfo{author}{Abanin, D.} \emph{et~al.}
\newblock \bibinfo{journal}{\bibinfo{title}{Distinguishing localization from
  chaos: Challenges in finite-size systems}}.
\newblock {\emph{\JournalTitle{Annals of Physics}}}
  \textbf{\bibinfo{volume}{427}}, \bibinfo{pages}{168415},
  \doiprefix\url{https://doi.org/10.1016/j.aop.2021.168415}
  (\bibinfo{year}{2021}).

\bibitem{variance_Sn/2}
\bibinfo{author}{Piemontese, S.}, \bibinfo{author}{Roscilde, T.} \&
  \bibinfo{author}{Hamma, A.}
\newblock \bibinfo{journal}{\bibinfo{title}{Entanglement complexity of the
  rokhsar-kivelson-sign wavefunctions}}.
\newblock {\emph{\JournalTitle{Physical Review B}}}
  \textbf{\bibinfo{volume}{107}}, \bibinfo{pages}{134202},
  \doiprefix\url{https://doi.org/10.1103/PhysRevB.107.134202}
  (\bibinfo{year}{2023}).

\bibitem{PRLwang}
\bibinfo{author}{Wang, Y.}, \bibinfo{author}{Cheng, C.}, \bibinfo{author}{Liu,
  X.-J.} \& \bibinfo{author}{Yu, D.}
\newblock \bibinfo{journal}{\bibinfo{title}{Many-body critical phase: Extended
  and nonthermal}}.
\newblock {\emph{\JournalTitle{Phys. Rev. Lett.}}}
  \textbf{\bibinfo{volume}{126}}, \bibinfo{pages}{080602},
  \doiprefix\url{https://doi.org/10.1103/PhysRevLett.126.080602}
  (\bibinfo{year}{2021}).

\bibitem{Styliaris2019}
\bibinfo{author}{Styliaris, G.}, \bibinfo{author}{Anand, N.},
  \bibinfo{author}{Campos~Venuti, L.} \& \bibinfo{author}{Zanardi, P.}
\newblock \bibinfo{journal}{\bibinfo{title}{Quantum coherence and the
  localization transition}}.
\newblock {\emph{\JournalTitle{Phys. Rev. B}}} \textbf{\bibinfo{volume}{100}},
  \bibinfo{pages}{224204},
  \doiprefix\url{https://doi.org/10.1103/PhysRevB.100.224204}
  (\bibinfo{year}{2019}).

\bibitem{Jethin2023}
\bibinfo{author}{Pulikkottil, J.~J.} \emph{et~al.}
\newblock \bibinfo{journal}{\bibinfo{title}{Quantum coherence controls the
  nature of equilibration and thermalization in coupled chaotic systems}}.
\newblock {\emph{\JournalTitle{Phys. Rev. E}}} \textbf{\bibinfo{volume}{107}},
  \bibinfo{pages}{024124},
  \doiprefix\url{https://doi.org/10.1103/PhysRevE.107.024124}
  (\bibinfo{year}{2023}).

\bibitem{Mitra2023}
\bibinfo{author}{Mitra, A.} \& \bibinfo{author}{Srivastava, S. C.~L.}
\newblock \bibinfo{journal}{\bibinfo{title}{Sunburst quantum ising model under
  interaction quench: Entanglement and role of initial state coherence}}.
\newblock {\emph{\JournalTitle{Phys. Rev. E}}} \textbf{\bibinfo{volume}{108}},
  \bibinfo{pages}{054114},
  \doiprefix\url{https://doi.org/10.1103/PhysRevE.108.054114}
  (\bibinfo{year}{2023}).

\bibitem{manju25}
\bibinfo{author}{Manju, C.}, \bibinfo{author}{Divakaran, U.} \&
  \bibinfo{author}{Lakshminarayan, A.}
\newblock \bibinfo{journal}{\bibinfo{title}{Disordering a permutation symmetric
  system: revivals, thermalisation and chaos}}.
\newblock {\emph{\JournalTitle{arXiv.2505.24453}}}
  \doiprefix\url{https://doi.org/10.48550/arXiv.2505.24453}
  (\bibinfo{year}{2025}).

\bibitem{RC_review2023}
\bibinfo{author}{Fisher, M.~P.}, \bibinfo{author}{Khemani, V.},
  \bibinfo{author}{Nahum, A.} \& \bibinfo{author}{Vijay, S.}
\newblock \bibinfo{journal}{\bibinfo{title}{Random quantum circuits}}.
\newblock {\emph{\JournalTitle{Annual Review of Condensed Matter Physics}}}
  \textbf{\bibinfo{volume}{14}}, \bibinfo{pages}{335--379},
  \doiprefix\url{https://doi.org/10.1146/annurev-conmatphys-031720-030658}
  (\bibinfo{year}{2023}).

\bibitem{MBL_Floquet2016}
\bibinfo{author}{Zhang, L.}, \bibinfo{author}{Khemani, V.} \&
  \bibinfo{author}{Huse, D.~A.}
\newblock \bibinfo{journal}{\bibinfo{title}{A floquet model for the many-body
  localization transition}}.
\newblock {\emph{\JournalTitle{Phys. Rev. B}}} \textbf{\bibinfo{volume}{94}},
  \bibinfo{pages}{224202},
  \doiprefix\url{https://doi.org/10.1103/PhysRevB.94.224202}
  (\bibinfo{year}{2016}).

\bibitem{MBL_Floquet2023}
\bibinfo{author}{Sierant, P.}, \bibinfo{author}{Lewenstein, M.},
  \bibinfo{author}{Scardicchio, A.} \& \bibinfo{author}{Zakrzewski, J.}
\newblock \bibinfo{journal}{\bibinfo{title}{Stability of many-body localization
  in floquet systems}}.
\newblock {\emph{\JournalTitle{Phys. Rev. B}}} \textbf{\bibinfo{volume}{107}},
  \bibinfo{pages}{115132},
  \doiprefix\url{https://doi.org/10.1103/PhysRevB.107.115132}
  (\bibinfo{year}{2023}).

\bibitem{trapped_ion_RevMod}
\bibinfo{author}{Monroe, C.} \emph{et~al.}
\newblock \bibinfo{journal}{\bibinfo{title}{Programmable quantum simulations of
  spin systems with trapped ions}}.
\newblock {\emph{\JournalTitle{Rev. Mod. Phys.}}}
  \textbf{\bibinfo{volume}{93}}, \bibinfo{pages}{025001},
  \doiprefix\url{https://link.aps.org/doi/10.1103/RevModPhys.93.025001}
  (\bibinfo{year}{2021}).

\bibitem{quantum_ion_disordered}
\bibinfo{author}{Trautmann, N.} \& \bibinfo{author}{Hauke, P.}
\newblock \bibinfo{journal}{\bibinfo{title}{Trapped-ion quantum simulation of
  excitation transport: Disordered, noisy, and long-range connected quantum
  networks}}.
\newblock {\emph{\JournalTitle{Phys. Rev. A}}} \textbf{\bibinfo{volume}{97}},
  \bibinfo{pages}{023606},
  \doiprefix\url{https://link.aps.org/doi/10.1103/PhysRevA.97.023606}
  (\bibinfo{year}{2018}).

\bibitem{MBL_disorder_exp}
\bibinfo{author}{Smith, J.} \emph{et~al.}
\newblock \bibinfo{journal}{\bibinfo{title}{Many-body localization in a quantum
  simulator with programmable random disorder}}.
\newblock {\emph{\JournalTitle{Nature Physics}}} \textbf{\bibinfo{volume}{12}},
  \doiprefix\url{https://doi.org/10.1038/nphys3783} (\bibinfo{year}{2015}).

\bibitem{Rydberg_exp}
\bibinfo{author}{Steinert, L.-M.} \emph{et~al.}
\newblock \bibinfo{journal}{\bibinfo{title}{Spatially tunable spin interactions
  in neutral atom arrays}}.
\newblock {\emph{\JournalTitle{Phys. Rev. Lett.}}}
  \textbf{\bibinfo{volume}{130}}, \bibinfo{pages}{243001},
  \doiprefix\url{https://link.aps.org/doi/10.1103/PhysRevLett.130.243001}
  (\bibinfo{year}{2023}).

\bibitem{sauerwein23}
\bibinfo{author}{Sauerwein, N.} \emph{et~al.}
\newblock \bibinfo{journal}{\bibinfo{title}{Engineering random spin models with
  atoms in a high-finesse cavity}}.
\newblock {\emph{\JournalTitle{Nature Physics}}} \textbf{\bibinfo{volume}{19}},
  \bibinfo{pages}{1128},
  \doiprefix\url{https://doi.org/10.1038/s41567-023-02033-3}
  (\bibinfo{year}{2023}).

\bibitem{russomanno2021quantum}
\bibinfo{author}{Russomanno, A.}, \bibinfo{author}{Fava, M.} \&
  \bibinfo{author}{Heyl, M.}
\newblock \bibinfo{journal}{\bibinfo{title}{Quantum chaos and ensemble
  inequivalence of quantum long-range ising chains}}.
\newblock {\emph{\JournalTitle{Physical Review B}}}
  \textbf{\bibinfo{volume}{104}}, \bibinfo{pages}{094309},
  \doiprefix\url{https://doi.org/10.1103/PhysRevB.104.094309}
  (\bibinfo{year}{2021}).

\end{thebibliography}

\section*{Acknowledgements}
 MC and UD acknowledge the HPC facility Chandra at IIT Palakkad where the computations were carried out. AL thanks Sylvia Pappalardi for discussions at an early stage of this work, at the program on ``Dynamical Foundations of Many-Body Chaos", Institut Pascal, Universit\'e Paris-Saclay, during March-April 2023, whose organizers are also gratefully acknowledged. 

\section*{Author Contributions}
All authors contributed equally to the analysis, interpretation of the results and the preparation of
the manuscript. 

\section*{Additional Information}
Supplementary material available. \\
Competing financial interests: The authors declare no competing financial interests.

\end{document}